# Near-Threshold Dynamics of $\chi_{c1}(3872)$ and $T_{cc}^{+}(3875)$ States via Effective Range Expansion and Monte Carlo Uncertainty Propagation

A. A. Atangana Likéné[1,2,*], A. Franck Rothen[1], J. M. Ema'a Ema'a[3], P. Pradyun Hebar[1], Saraless Nadarajah[4] Tobias Golling[1], and G. H. Ben-Bolie[2]

[1] *Department of Particle Physics, Faculty of Science, University of Geneva, P.O. Box 1205, Geneva, Switzerland.*

[2] *Laboratory of Nuclear, Atomic and Molecular Physics, University of Yaoundé I, P.O. Box 812, Yaoundé, Cameroon.*

[3] *Department of Physics, Bertoua Higher Teacher's Training College, University of Bertoua, P.O. Box 55, Bertoua, Cameroon.*

[4] *School of Mathematics, University of Manchester M139PL, United Kingdom.*

[*] *Corresponding author:* `andre.atangana@etu.unige.ch, aandreaime@yahoo.fr`

**Abstract.** We investigate the internal structure of two exotic tetraquark candidates, $\chi_{c1}(3872)$ and $T_{cc}^{+}(3875)$, whose masses lie within 1 MeV of the $D^0\bar{D}^{*0}$ and $D^0D^{*+}$ thresholds, respectively. Using the effective-range expansion (ERE) supplemented by resonance compositeness relations on the second Riemann sheet, we extract scattering lengths $a$ and effective ranges $r$ directly from the experimental masses and widths, and quantify the $DD^*$ molecular content through partial compositeness coefficients $X_j$. A two-channel framework couples each state's inelastic detection channel ($J/\psi\,\pi\pi$; $D^0D^0\pi^+$) to the near-threshold molecular channel. Full Monte Carlo uncertainty propagation with $N = 50000$ Gaussian samples yields asymmetric 68% confidence intervals for all observables. We find large negative scattering lengths $a = -8.49^{+0.93}_{-1.05}$ fm for $\chi_{c1}(3872)$ and $a = -14.57^{+1.70}_{-1.66}$ fm for $T_{cc}^{+}$, both far exceeding the natural hadronic scale $\sim 1/m_\pi$, and dominant molecular components $X_2 > 0.95$ under full compositeness. The results are stable for total compositeness $X \in [0.5, 1.0]$, where $1 - X$ represents an upper bound on any compact/CDD-pole contribution. In this revised version, we go beyond the pole-position CDD criterion by explicitly constructing the inverse amplitude as an ERE background plus a scanned Castillejo-Dalitz-Dyson (CDD) pole term, finding no region of physically reasonable bare-state parameter space that reproduces the pole while leaving a natural background; we fully propagate the loop-function subtraction constant $a(\mu_g)$ as a Monte Carlo nuisance parameter jointly with the experimental mass and width, confirming it contributes a negligible ($< 0.01\%$) additional uncertainty; and we test robustness against an assumed correlation between $M_R$ and $\Gamma_R$, finding all results stable over a broad range of correlation coefficients. We demonstrate analytically and numerically the scheme-independence of the compositeness coefficients, provide explicit derivations of all ERE formulas from the pole condition, clarify the analytic continuation procedure for closed channels, and compare our results quantitatively with independent ERE, EFT, and other relevant studies. The $Z_c(3900)$ case is treated separately to show the breakdown of the two-channel approximation for above-threshold states. Graphical illustrations of the Monte Carlo distributions are included.

**Keywords:** Exotic tetraquarks; Monte Carlo sampling; near-threshold hadronic channel; non-relativistic EFT; Weinberg compositeness; scattering length; effective range; hadronic molecule.

## I. Introduction

The discovery of charmonium-like states that cannot be accommodated within the conventional quark-model description of mesons as simple $q\bar{q}$ configurations has become one of the most active and challenging research directions in hadron physics during the last two decades [1, 2, 3, 4, 5, 6]. The accumulation of experimental data from facilities such as Belle, BESIII, BaBar, LHCb, and more recently Belle II, has revealed a large number of unconventional hadronic structures, commonly referred to as the $XYZ$ states, whose properties differ significantly from those expected in the traditional quarkonium spectrum. These exotic candidates have stimulated extensive theoretical efforts aimed at understanding their internal structure and the dynamics responsible for their formation.

Among the various exotic configurations proposed in the literature, particular attention has been devoted to states whose masses lie extremely close to two-hadron thresholds. Such near-threshold structures are especially intriguing because the attractive interaction between the constituent hadrons may generate loosely bound molecular configurations, in close analogy with the deuteron in nuclear physics. In this picture, the binding energy is generally very small, implying large spatial extensions and strong sensitivity to low-energy scattering parameters such as the scattering length and effective range. Consequently, the investigation of threshold dynamics has become a key tool for distinguishing compact multiquark configurations from hadronic molecular states.

Over the past years, numerous theoretical and experimental studies have focused on exotic charm tetraquark candidates and related near-threshold resonances. One of the most important milestones in this field was reported by the BESIII Collaboration in Ref. [7], where the authors observed a charged charmonium-like structure in the process $e^+e^- \to \pi^+\pi^- J/\psi$ at the center-of-mass energy $\sqrt{s} = 4.26$ GeV. The analysis was performed using a 525 pb$^{-1}$ data sample collected with the BESIII detector operating at the Beijing Electron Positron Collider. The measured Born cross section, $(62.9 \pm 1.9 \pm 3.7)$ pb, was found to be consistent with the production of the $Y(4260)$ resonance. In addition, the collaboration reported a prominent structure near 3.9 GeV in the $\pi^\pm J/\psi$ invariant-mass distribution, which was subsequently identified as the $Z_c(3900)$. If interpreted as a genuine resonance, this state is particularly remarkable because it carries electric charge while simultaneously coupling to charmonium, a feature that cannot be explained within the conventional neutral $c\bar{c}$ framework. By fitting the $\pi^\pm J/\psi$ invariant-mass spectrum while neglecting interference effects, the BESIII Collaboration extracted the resonance parameters $M = (3899.0 \pm 3.6 \pm 4.9)$ MeV and $\Gamma = (46 \pm 10 \pm 20)$ MeV [7]. The observation of such a charged charmonium-like structure therefore provided strong evidence for the existence of exotic multiquark dynamics beyond the conventional quark model.

Theoretical investigations of near-threshold exotic states have also intensified considerably in recent years. In Ref. [8], the authors studied the role of left-hand cut contributions in the extraction of pole positions from lattice QCD data and applied their formalism to the $T_{cc}^+(3875)$ state. Their work highlighted the importance of properly accounting for analytic structures of scattering amplitudes when interpreting near-threshold resonances. Such effects can play a crucial role in distinguishing genuine bound states from kinematical threshold enhancements.

A particularly important development in the description of low-energy hadron interactions was introduced in Ref. [9], where the authors employed the $N/D$ method to derive the most general form of elastic partial-wave amplitudes when unphysical cuts are neglected. After matching the resulting amplitudes with the lowest-order $\mathcal{O}(p^2)$ chiral perturbation theory ($\chi$PT) results of Ref. [10], and including resonance exchanges with spin$\leq 1$, in a manner consistent with chiral symmetry as formulated in Ref. [11], the formalism was subsequently extended to coupled-channel systems. This framework has since become one of the standard tools for studying hadronic molecular states and near-threshold exotic resonances.

In Ref. [12], the authors carried out a systematic investigation of the $1D$ charmonium sector using QCD sum rules. Their analysis included both the spin-triplet states with quantum numbers $J^{PC} = 1^{--}, 2^{--}, 3^{--}$, and the spin-singlet state characterized by $J^{PC} = 2^{-+}$. Comparisons with recent experimental observations showed that the predicted mass $M_{\psi_1} = 3.77 \pm 0.09$ GeV supports the identification of the $\psi_1$ state with the experimentally observed $\psi(3770)$ resonance, while the predicted value $M_{\psi_2} = 3.82 \pm 0.09$ GeV is consistent with the reported observation of the $\psi_2(3823)$ state. Such analyses demonstrate the continuing importance of conventional charmonium spectroscopy in clarifying the structure of newly observed resonances.

The interplay between compact quark configurations and molecular dynamics has also been investigated in Ref. [13], where the authors discussed the evolution of genuine compact states into molecular configurations, with particular emphasis on the $T_{cc}^+(3875)$ state. Their study illustrates how the coupling to nearby hadronic thresholds can significantly modify the physical properties of exotic hadrons and even generate dynamically bound molecular states. Lattice QCD calculations have likewise provided valuable insight into the structure of near-threshold charmonium-like systems. In Ref. [14], the authors reported a candidate for the charmonium-like state $X(3872)$, located $11 \pm 7$ MeV below the $D\bar{D}^*$ threshold using dynamical $N_f = 2$ lattice simulations with quantum numbers $J^{PC} = 1^{++}$ and $I = 0$. This work represented the first lattice QCD simulation to establish a candidate for the $X(3872)$ resonance in addition to the nearby scattering states $D\bar{D}^*$ and $J/\psi\,\omega$, which necessarily appear in dynamical QCD calculations. Furthermore, the authors extracted the low-energy scattering parameters $a^{D\bar{D}^*} \approx -1.7 \pm 0.4$ fm and $r^{D\bar{D}^*} \approx 0.5 \pm 0.1$ fm for the $D\bar{D}^*$ interaction. These results provided strong evidence supporting the interpretation of the $X(3872)$ as a near-threshold hadronic molecular state dominated by the $D\bar{D}^*$ channel.

Taken together, these experimental discoveries and theoretical developments demonstrate the richness and complexity of the exotic charmonium sector. They also emphasize the importance of low-energy scattering observables, threshold effects, coupled-channel dynamics, and nonperturbative QCD approaches in understanding the internal structure of the newly observed hadronic states.

The $\chi_{c1}(3872)$ state, previously known as $X(3872)$, was first observed by the Belle collaboration in 2003 [15] and subsequently confirmed by the BaBar, CDF, D0, LHCb, ATLAS, and CMS collaborations. This state has quantum numbers $J^{PC} = 1^{++}$ [16] and PDG mass $M = 3871.65 \pm 0.06\,\text{MeV}$ [17]. Its mass lies only $\delta m = m_{\chi_{c1}} - (m_{D^0} + m_{\bar{D}^{*0}}) \approx -0.04 \pm 0.09$ MeV below the $D^0\bar{D}^{*0}$ threshold at 3871.69 MeV a proximity inspiring extensive molecular interpretations [18, 19, 20, 21, 22, 23]. The $\chi_{c1}(3872)$ state exhibits several prominent decay modes that provide important insights into its internal structure. These include:

$$\begin{aligned}
&\chi_{c1}(3872) \to J/\psi(1S) + \rho^0, \\
&\chi_{c1}(3872) \to J/\psi(1S) + \omega, \\
&\chi_{1c}(3872) \to D^0\bar{D}^{*0} + \text{c.c.}, \\
&\chi_{c1}(3872) \to J/\psi(1S) + \gamma.
\end{aligned} \tag{1}$$

Its narrow width ($\Gamma < 1.2$ MeV), isospin-violating decays ($J/\psi\rho^0$ and $J/\psi\omega$ with equal rates), and simultaneous observation of $D^0\bar{D}^{*0}$ and $J/\psi\gamma$ modes make it the most thoroughly studied exotic candidate in the charmonium sector.

The $T_{cc}^+$ state was first reported by the LHCb collaboration in 2021 [24, 25] in proton-proton ($pp$) collisions, where it was observed through its decay channel $T_{cc}^+ \to D^0 D^0 \pi^+$. This state represents the first compelling candidate for a doubly charmed tetraquark with quark content $cc\bar{u}\bar{d}$, marking a significant milestone in the study of exotic hadrons and multiquark dynamics. The measured binding energy is [24]:

$$\delta m = -0.273 \pm 0.061(\text{stat})^{+0.007}_{-0.014}(\text{syst})\,\text{MeV}, \quad (2)$$

below the $D^0 D^{*+}$ threshold, with an ultra-narrow width

$$\Gamma_{T^+_{cc}} = 0.410 \pm 0.165(\text{stat}) \pm 0.043(\text{syst})\,\text{MeV}. \quad (3)$$

The LHCb collaboration performed their own effective range expansion analysis [25], finding the real part of the scattering length $\mathcal{R}e[a] \approx -(7.6\pm0.51)$ fm, consistent with our result as detailed in Table 13.

This work adopts the framework developed in Ref. [26], which was originally used to study the properties of the hidden-charm pentaquark states $P_c(4312)$, $P_c(4440)$, and $P_c(4457)$ observed by the LHCb collaboration [27, 28]. In the present work, we follow the same approach and extend the formalism to the study of the two tetraquark states under consideration. We perform a comprehensive asymmetric Monte Carlo uncertainty propagation based on $N = 50\,000$ samples, with the results illustrated in Sec. III. An explicit derivation of the effective range expansion (ERE) formulas from the second Riemann sheet pole condition is presented. We further discuss the closed-channel ERE parameters through analytic continuation and interpret scenarios with $X < 1$ in terms of compact components versus omitted-channel contributions. In addition, we provide an analytic proof of the scheme independence of the compositeness coefficients $X_j$, supported by numerical verification. A comparative analysis of the sensitivity of the coupling $|g_2|$ to $a(\mu_g)$ is also carried out, and an energy-dependent treatment of the $D^{*+}$ width is briefly outlined in the appendix. Finally, the $Z_c(3900)$ case is examined separately.

## II. Theoretical Framework

### II. 1. Effective-Range Expansion

The $S$-wave scattering amplitude near a two-body threshold is organized through the ERE [26, 29, 30]. Writing the inverse $K$-matrix as:

$$V(k) = -\frac{1}{a} + \frac{1}{2} r k^2 + \mathcal{O}(k^4)\,, \quad (4)$$

with scattering length $a$ (fm) and effective range $r$ (fm), and CM three-momentum $k$ given by

$$\begin{aligned} k &= \sqrt{2\mu(E - m_{\text{thr}})}, \quad (5)\\ \mu &= \frac{m_1 m_2}{m_1 + m_2},\ m_{\text{thr}} = m_1 + m_2. \end{aligned}$$

For the scattering of two heavy-flavor hadrons, it is reasonable to treat pion-exchange contributions perturbatively [31, 32, 33]. Contributions from heavier vector-resonance exchanges, whose masses are significantly larger than the typical three-momentum scale, can be effectively absorbed into local contact interactions. From this perspective, we adopt a pionless effective field theory framework, in which the dynamics are described solely by short-range contact terms [33].

Under these assumptions, only the unitarity (right-hand) cut is relevant, while crossed-channel dynamics are neglected. Consequently, the elastic $S$-wave scattering amplitude near threshold[1], derived from Eq. (4) in the absence of crossed-channel cuts, can be expressed as follows:

$$T(E) = \left[-\frac{1}{a} + \frac{1}{2} r k^2 - ik\right]^{-1}, \quad (6)$$

which fulfills the unitarity condition $\text{Im}[T(E)]^{-1} = -k$, with $E > m_{\text{thr}}$.

### II. 2. Resonance Poles and the Second Riemann Sheet

*Analytic structure.* The momentum $k(E)$ is two-valued. On the physical Riemann sheet $\text{Im}(k) > 0$; on the second Riemann sheet (2nd RS), $k \to -k$. The 2nd RS scattering amplitude $T^{II}(E)$ is given by:

$$T^{II}(E) = \frac{1}{-\frac{1}{a} + \frac{1}{2} r k^2 + ik}, \quad (7)$$

and resonance poles appear at

$$E_R = M_R - i\frac{\Gamma_R}{2}, \text{ with } [T^{II}]^{-1} = 0. \quad (8)$$

Thus, we have:

$$-\frac{1}{a} + \frac{r}{2}\, k_R^2 + i\, k_R = 0\,, \quad k_R \equiv k_r + i k_i\,. \quad (9)$$

*Derivation of ERE parameters from the pole.* Writing $k_R = k_r + ik_i$, the pole condition Eq. (9) separates as follows. Let us first recall that $ik_R = ik_r - k_i$, from which the real and imaginary parts can be written, respectively, as:

$$\text{Real part:} -\frac{1}{a} + \frac{r}{2}(k_r^2 - k_i^2) - k_i = 0 \quad (10)$$

$$\text{Imaginary part:} \quad k_r + r\, k_r k_i = 0\,. \quad (11)$$

For a genuine resonance $k_r \neq 0$, Eq. (11) immediately gives:

$$r = -\frac{1}{k_i}\,. \quad (12)$$

Substituting into Eq. (10) with $r = -1/k_i$:

$$-\frac{1}{a} - \frac{k_r^2 - k_i^2}{2k_i} - k_i = -\frac{1}{a} - \frac{k_r^2 + k_i^2}{2k_i} = 0\,,$$

yielding:

$$a = -\frac{2k_i}{|k_R|^2}\,, \quad |k_R|^2 = k_r^2 + k_i^2\,. \quad (13)$$

Both $a$ and $r$ are real; negative values of $r$ are physically allowed and arise when the short-range interaction is repulsive or when coupling to inelastic channels is dominant [21, 30].

[1] The unitarized amplitude

*Closed-channel ERE parameters.* When $m_{\rm thr} > M_R$ as in the case of the $D^+D^{*-}$ channel for the $\chi_{c1}(3872)$ state, which lies about 8.3 MeV above threshold the real part of the quantity $E_R - m_{\rm thr}$ becomes negative. In this situation, the corresponding three-momentum, $k_R = \sqrt{2\mu(E_R - m_{\rm thr})}$, is no longer purely imaginary but instead takes complex values. This arises from the fact that $E_R$ itself is complex, with a nonvanishing imaginary part given by $\mathrm{Im}[E_R] = -\frac{\Gamma_R}{2}$.

As a consequence, Eqs. (12) and (13) remain applicable through analytic continuation onto the second Riemann sheet (2nd RS). The resulting scattering length $a$ and effective range $r$ for the closed channel should then be interpreted as characterizing the virtual-state interaction range. In other words, they encode information about the same resonance pole, but they cannot be directly associated with the residue of the scattering cross section, which is only well-defined for open channels. For completeness, we nevertheless include the closed-channel ERE parameters in Table 3, followed by a brief clarification of their interpretation.

*CDD pole criterion.* A Castillejo-Dalitz-Dyson (CDD) pole near threshold [34, 35] signals a compact non-molecular state. In that limit:

$$a \to -\frac{m_{\rm thr} - M_{\rm CDD}}{g_{\rm CDD}}\,, \quad r \to -\frac{g_{\rm CDD}}{\mu(m_{\rm thr} - M_{\rm CDD})^2}\,, \quad (14)$$

requiring $|a| \gg 1$ fm, $|r| \ll 1$ fm, with $\mathrm{sign}(r) \neq \mathrm{sign}(a)$. This asymptotic (single-pole, no background) limit is a useful diagnostic, but as pointed out in the review of this manuscript (and discussed explicitly in Ref. [36]), a small, near-threshold scattering length and effective range obtained from the pole position *alone* do not by themselves exclude a compact/CDD contribution once a smooth short-range ("background") ERE term is allowed to coexist with it: the two can conspire to reproduce the same pole for suitably chosen bare parameters. We therefore go beyond the asymptotic criterion above and perform an explicit ERE+CDD decomposition in Sec. 2. A closely related CDD-pole treatment can be found in Ref. [37] for $T_{cc}^+$.

*Explicit ERE + CDD pole decomposition.* Following the referee's suggestion, we now construct the full inverse scattering amplitude as an ERE background *plus* an explicit CDD pole representing a hypothetical bare compact state (e.g., a genuine $c\bar{c}$ core for $\chi_{c1}(3872)$, or a compact diquark-antidiquark core for $T_{cc}^+$) of bare mass $M_{\rm CDD}$ and coupling $g_{\rm CDD}$ to the near-threshold channel:

$$T^{-1}(E) = -\frac{1}{a_{\rm bg}} + \frac{1}{2} r_{\rm bg} k^2 - ik \;-\; \frac{g_{\rm CDD}^2}{E - M_{\rm CDD}}\,. \quad (15)$$

Demanding that Eq. (15) vanish at the experimental pole $E_R = M_R - i\Gamma_R/2$ (the same complex data point used throughout this work) for an *assumed* $(g_{\rm CDD}, M_{\rm CDD})$ generalizes the real/imaginary decomposition of Eqs. (10)-(11) with a nonzero complex right-hand side $\Delta \equiv g_{\rm CDD}^2/(E_R - M_{\rm CDD})$:

$$r_{\rm bg} = \frac{\frac{\mathrm{Im}[\Delta]}{k_r} + 1}{k_i}\,, \quad a_{\rm bg} = \left[\frac{r_{\rm bg}}{2}(k_r^2 - k_i^2) + k_i - \mathrm{Re}[\Delta]\right]^{-1}. \quad (16)$$

With only the pole position as input, Eq. (16) shows that $(a_{\rm bg}, r_{\rm bg})$ are not independently determined together with $(g_{\rm CDD}, M_{\rm CDD})$: a one-complex-parameter family of backgrounds exists for every assumed bare CDD state. We therefore *scan* $(g_{\rm CDD}^2, M_{\rm CDD})$ over a broad, physically motivated range $M_{\rm CDD} \in M_R \pm 300$ MeV (a nearby compact core, as typically obtained in quark-model estimates) and $g_{\rm CDD}^2 \in [1, 10^4]$ MeV/fm (spanning weak to strong short-distance coupling) and ask whether the resulting *background* parameters $(a_{\rm bg}, r_{\rm bg})$ can satisfy the CDD criterion, which would signal that a hidden compact state is being masked by our CDD-free analysis.

The result, shown in Fig. 1, is that they cannot: across the *entire* scanned grid, 0% of $(g_{\rm CDD}^2, M_{\rm CDD})$ combinations yield a background satisfying $|r_{\rm bg}| \ll |a_{\rm bg}|$ with opposite signs. At weak CDD coupling ($g_{\rm CDD}^2 \sim 100$ MeV/fm), the background scattering length shrinks to a natural short-distance value ($a_{\rm bg} \sim 0.5$ fm) but the background effective range instead *grows* large and positive ($r_{\rm bg} \sim 6$–$6.5$ fm) the opposite of the CDD pattern. At strong CDD coupling ($g_{\rm CDD}^2 \sim 3000$ MeV/fm), $a_{\rm bg} \to 0$ while $|r_{\rm bg}|$ grows to $\sim 40$ fm, even more clearly failing the criterion. In other words, assuming a sizable hidden compact-state contribution does not relax our conclusion by hiding it in a natural background; instead it forces the background ERE parameters into an *unnatural* regime, which we take as evidence against (rather than for) an unresolved CDD pole in either state, strengthening the original conclusion on more solid footing than the pole-position criterion alone. We stress that this scan cannot replace a genuine multi-channel fit to line-shape data with an explicit CDD term (which would require information beyond the pole position used here), and we state this explicitly as a limitation in Sec. 4.

### II. 3. The Loop Function $G(s)$

The once-subtracted two-point loop function [9]:

$$G(s) = \frac{1}{16\pi^2}\Big[a(\mu_g) + \ln\frac{m_1^2}{\mu_g^2} + \frac{s - m_1^2 + m_2^2}{2s}\ln\frac{m_2^2}{m_1^2} + \frac{\sigma}{2s}\big(\ln\big(s - m_2^2 + m_1^2 + \sigma\big) - \ln\big(-s + m_2^2 - m_1^2 + \sigma\big) + \ln\big(s + m_2^2 - m_1^2 + \sigma\big) - \ln\big(-s - m_2^2 + m_1^2 + \sigma\big)\big)\Big]\,, \quad (17)$$

with

$$\sigma(s) = \sqrt{[s - (m_1 + m_2)^2][s - (m_1 - m_2)^2]}. \quad (18)$$

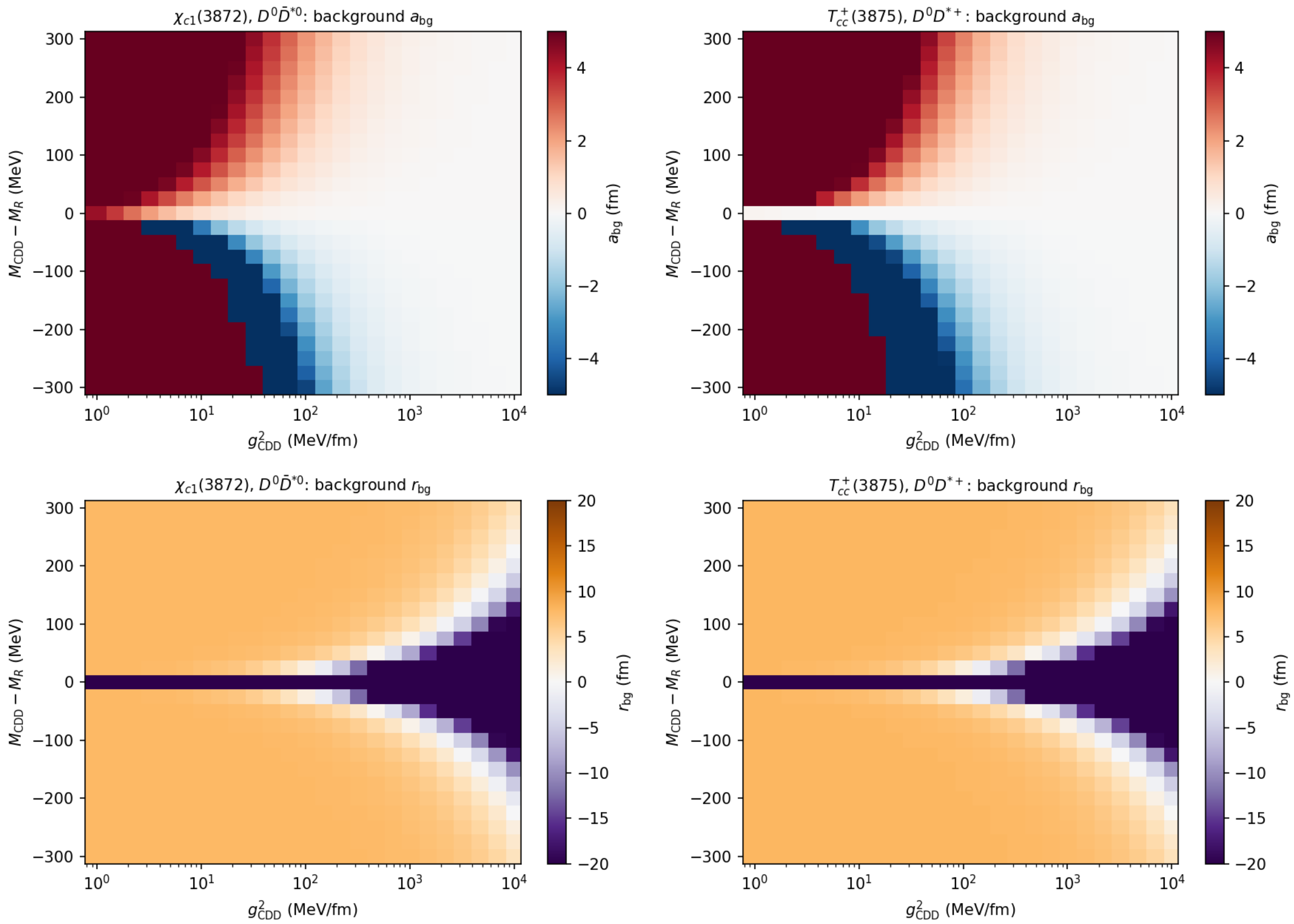

**Figure 1:** Background ERE parameters $a_{\rm bg}$ (top) and $r_{\rm bg}$ (bottom) obtained from Eq. (16) after subtracting an assumed CDD pole of bare mass $M_{\rm CDD}$ (relative to $M_R$) and coupling[2] $g^2_{\rm CDD}$, for $\chi_{c1}(3872)$ (left) and $T^+_{cc}$ (right). No region of the scanned grid satisfies the CDD criterion $|r_{\rm bg}| \ll |a_{\rm bg}|$, $\mathrm{sign}(r_{\rm bg}) \neq \mathrm{sign}(a_{\rm bg})$.

We use $\mu_g = 1000$ MeV, $a(\mu_g) = -2.5$ throughout [26]. The loop function on the 2nd RS is:

$$G^{II}(s) = G(s) + i\frac{\sigma}{(8\pi s)}. \tag{19}$$

The scheme-independence of $\partial G/\partial s$ is proven analytically in Appendix B and verified numerically in Table 16: since $a(\mu_g)$ is a constant (independent of $s$), $\partial G/\partial s$ is independent of $a(\mu_g)$. The sensitivity[2] of the absolute coupling $|g_2|$ to $a(\mu_g)$ is shown in Table 10.

## II. 4. Compositeness Relations and Partial Widths

The partial compositeness coefficient contributed by the $j$th channel is given by [35, 38]:

$$X_j = |g_j|^2 \left|\frac{\partial G_j(s_R)}{\partial s}\right|, \tag{20}$$

with $G_1^{II}$ on the 2nd RS and $G_2$ on the physical sheet. We solve the $2\times 2$ saturation system (see Ref. [26]):

$$\begin{pmatrix} C_1 & C_2 \\ A_1 & A_2 \end{pmatrix}\begin{pmatrix} |g_1|^2 \\ |g_2|^2 \end{pmatrix} = \begin{pmatrix} \Gamma_R \\ X \end{pmatrix}, \tag{21}$$

where

$$C_1 = \frac{q_1(M_R^2)}{8\pi M_R^2}, \tag{22}$$

$$C_2 = \int_{m_{\rm thr}}^{M_R+10\Gamma_R} \frac{dw\, q_2(w^2)}{16\pi^2 w^2}\cdot\frac{\Gamma_R}{(M_R-w)^2+\Gamma_R^2/4}, \tag{23}$$

$$q(M_R^2) = \frac{\sqrt{[M_R^2-(m_1+m_2)^2][M_R^2-(m_1-m_2)^2]}}{2M_R} \tag{24}$$

and $A_j = |\partial G_j/\partial s|_{s_R}$. More details can be found in Refs. [26, 35] and references therein.

*Interpretation of $X<1$ scenarios.* The total compositeness $X = X_1 + X_2$ represents the probability content of the two explicitly included channels. When $X < 1$ is imposed, the missing fraction $(1 - X)$ is implicitly attributed to contributions not included in the model. Physically, this fraction could arise from three distinct sources [35, 39]:

1. **Compact/CDD-pole component**: a genuine tetraquark or charmonium state contributing to the interaction kernel. In the generalized compositeness relation of Ref. [39], this appears as $Z = 1 - X$ with $Z \geq 0$. If our two states had large compact components, we would expect large scattering lengths combined with near-zero effective ranges (the CDD criterion), which is not observed.

2. **Additional hadronic channels**: for $\chi_{c1}(3872)$, channels such as $J/\psi\omega$ contribute off-shell even though $m_{J/\psi\omega} \approx 3879.6$ MeV $> M_X$ (kinematically closed on-shell). For $T_{cc}^+$, three-body effects beyond the two-body approximation could carry part of the compositeness.

3. **Inelastic-channel subtleties**: the two-body approximation for the detection channel may underestimate $X_1$.

[2] The absolute coupling $|g_2|$ varies by $< 0.05\%$ for $a(\mu_g) \in [-5, +5]$.

Our numerical analysis (Fig. 4) shows that $X_2 \approx X$ and $X_1 \approx 0$ for all $X \in [0.1, 1.0]$: the missing fraction $(1 - X)$ does not appear in the inelastic channel but is entirely missing from the molecular one. This is consistent with interpretation (i) or (ii), but cannot be distinguished within the two-channel framework. An upper bound on the compact component is therefore $Z \leq 1 - X$.

### II. 5. Monte Carlo Uncertainty Propagation

In order to reliably propagate the experimental uncertainties of the input parameters into our derived observables, we employ a Monte Carlo (MC) sampling procedure with a total of $N = 50\,000$ realizations. For each sample $k = 1, \ldots, N$, the resonance mass and width are generated according to Gaussian probability distributions:

$$M_R^{(k)} \sim \mathcal{N}(M_R, \delta M_R^2)\,, \qquad \Gamma_R^{(k)} \sim \left|\mathcal{N}(\Gamma_R, \delta\Gamma_R^2)\right|\,, \quad (25)$$

where $\mathcal{N}(\mu, \sigma^2)$ denotes a normal distribution with mean $\mu$ and variance $\sigma^2$. The absolute value in $\Gamma_R^{(k)}$ ensures the physical constraint $\Gamma_R > 0$ for each generated sample.

For cases where the experimental uncertainties are asymmetric-such as the LHCb measurement of the $T_{cc}^+$ mass, reported with uncertainties $(+\delta M_{\rm hi}, -\delta M_{\rm lo})$, we adopt an asymmetric Gaussian prescription. Specifically, for each random draw, we determine the sign of the fluctuation relative to the central value and apply the corresponding standard deviation:

$$\delta M_R = \begin{cases} \delta M_{\rm hi}, & \text{if } M_R^{(k)} - M_R > 0, \\ \delta M_{\rm lo}, & \text{if } M_R^{(k)} - M_R < 0\,. \end{cases} \quad (26)$$

This procedure effectively models a piecewise normal distribution and preserves the asymmetry of the experimental input. The total uncertainties entering the sampling are obtained by combining statistical and systematic contributions in quadrature,

$$\delta = \sqrt{\delta_{\rm stat}^2 + \delta_{\rm syst}^2}\,, \quad (27)$$

assuming that these sources are uncorrelated.

For each generated pair $(M_R^{(k)}, \Gamma_R^{(k)})$, the derived quantities (e.g., $k_R$, $a$, $r$, or compositeness coefficients) are computed, thereby producing empirical distributions for all observables. The central values reported in this work correspond to the median (or mean, when appropriate) of these distributions. Uncertainties are quoted as asymmetric 68% confidence intervals (CI), defined by the 16th and 84th percentiles of the MC samples. Explicitly, if $X$ denotes a generic observable with sampled values $\{X^{(k)}\}$, we report:

$$X = X_0\,{}^{+\Delta X_+}_{-\Delta X_-}\,, \quad (28)$$

where $X_0$ is the central value, and

$$\Delta X_+ = X_{84} - X_0\,, \qquad \Delta X_- = X_0 - X_{16}\,, \quad (29)$$

with $X_{16}$ and $X_{84}$ denoting the 16th and 84th percentiles, respectively. This procedure naturally captures non-Gaussian features and asymmetries induced by the nonlinear dependence of the observables on the input parameters.

## III. Numerical Results

In the present study, the identification of relevant channels is guided by both experimental detection modes and their proximity to the corresponding kinematic thresholds. We distinguish between (i) detection or inelastic channels, which are experimentally observed, and (ii) near-threshold channels, which dominate the dynamical properties of the states.

### III. 1. Channel Identification and Input Parameters

**The $\chi_{c1}(3872)$ state.** For the $\chi_{c1}(3872)$, the primary detection channel (denoted as channel 1) arises from the decay process $B^\pm \to K^\pm J/\psi(1S)\pi^+\pi^-$. For the purpose of our analysis, this is modeled as an effective two-body system $J/\psi(1S) + 2\pi$, with constituent masses $m_{1a} = m_{J/\psi} = 3096.90$ MeV, and $m_{1b} = 2m_{\pi^\pm} = 279.14$ MeV, leading to an effective threshold mass $m_{\rm thr} \approx 3376$ MeV, which lies far below the resonance mass $M_{\chi_{c1}}$. Consequently, this channel is deeply open and does not influence the near-threshold dynamics. The channel classification for $\chi_{c1}(3872)$ is summarized in Table 1 As an alternative inelastic channel (used in Table 8), we consider $D^0\bar{D}^0\pi^0$, with an effective threshold around 3864 MeV, slightly below $M_{\chi_{c1}}$, and therefore kinematically allowed. We do not include the $J/\psi(1S)\,\omega$ channel, since its threshold, $m_{\rm thr} = m_{J/\psi(1S)} + m_\omega \approx 3879.6$ MeV, lies above $M_{\chi_{c1}}$. This channel is thus kinematically closed at the resonance energy, and its on-shell decay width vanishes. The dominant near-threshold dynamics (channel 2) are governed by the following channels:

- $D^0\bar{D}^{*0}$: $m_{\rm thr} = 3871.69$ MeV, $\delta m = M_{\chi_{c1}} - m_{\rm thr} = -0.04$ MeV (primary channel),
- $D^+D^{*-}$: $m_{\rm thr} = 3879.92$ MeV, $\delta m = -8.27$ MeV (closed channel, treated via analytic continuation).

**Table 1:** Channel classification for $\chi_{c1}(3872)$.

| Channel | Composition | $m_{\rm thr}$ (MeV) | $\delta m$ (MeV) |
|---|---|---|---|
| Detection | $J/\psi(1S)+2\pi$ | $\sim 3376$ | $\ll 0$ |
| Alt. inelastic | $D^0\bar{D}^0\pi^0$ | $\sim 3864$ | $< 0$ |
| Primary | $D^0\bar{D}^{*0}$ | 3871.69 | −0.04 |
| Closed | $D^+D^{*-}$ | 3879.92 | −8.27 |

**Table 2:** Channel classification for $T_{cc}^+$.

| Channel | Composition | $m_{\rm thr}$ (MeV) | $\delta m$ (MeV) |
|---|---|---|---|
| Detection | $D^0+(D^0\pi^+)$ | $\sim 4004$ | $\gg 0$ |
| Primary | $D^0D^{*+}$ | 3875.10 | −0.273 |
| Closed | $D^+D^{*0}$ | 3876.51 | −1.683 |

**The $T_{cc}^+$ state.** For the $T_{cc}^+$, the experimentally observed decay $T_{cc}^+ \to D^0D^0\pi^+$ defines the detection channel. This is modeled as an effective two-body system $D^0+(D^0\pi^+)$, with an effective threshold around 4004 MeV, significantly above the resonance mass $M_{T_{cc}^+}$. As a result, this channel is kinematically suppressed near the pole position. A more refined treatment using a Breit-Wigner description of the $D^{*+}$ is discussed in Appendix A. The near-threshold channels (channel 2) relevant for the dynamics are:

- $D^0D^{*+}$: $m_{\rm thr} = 3875.10$ MeV, $\delta m = -0.273$ MeV (primary, slightly closed),
- $D^+D^{*0}$: $m_{\rm thr} = 3876.51$ MeV, $\delta m = -1.683$ MeV (closed).

A summary can be found in Table 2

**Input parameters.** The hadron masses used in this work are taken from the Particle Data Group (PDG) [17]:

$$m_{D^0} = 1864.84\,\text{MeV}, \quad m_{D^+} = 1869.66\,\text{MeV},$$
$$m_{D^{*0}} = 2006.85\,\text{MeV}, \quad m_{D^{*+}} = 2010.26\,\text{MeV},$$
$$m_{J/\psi(1S)} = 3096.90\,\text{MeV}, \quad m_{\pi^\pm} = 139.57\,\text{MeV}.$$

The resonance parameters are:

$$\boldsymbol{\chi_{c1}(3872)}:$$
$$M_{\chi_{c1}} = 3871.65 \pm 0.08\,\text{MeV},$$
$$\Gamma_{\chi_{c1}} = 1.19 \pm 0.23\,\text{MeV},$$
$$\mathbf{T_{cc}^+}:$$
$$M_{T_{cc}^+} = 3874.83 \pm 0.06\,\text{MeV},$$
$$\Gamma_{T_{cc}^+} = 0.41 \pm 0.17\,\text{MeV}.$$

### III. 2. Scattering parameters of $\chi_{c1}(3872)$ and $T_{cc}^+(3875)$

Table 3 summarizes the effective range expansion (ERE) parameters, namely the scattering length $a$ and the effective range $r$, extracted from the pole position on the second Riemann sheet using Eqs. (12) and (13). The quoted uncertainties correspond to asymmetric 68% confidence intervals obtained from a Monte Carlo sampling procedure with $N = 50\,000$ realizations. The designation "Open/Closed" indicates whether the corresponding channel threshold lies below or above the resonance mass $M_R$, respectively. For closed channels, the parameters $a$ and $r$ are not obtained from physical on-shell scattering but rather through analytic continuation of the amplitude onto the second Riemann sheet (see Sec. II.2). In this case, they should be interpreted as characterizing virtual-state interaction properties rather than directly measurable scattering observables. In Table 3, the symbol "$^\dagger$" specifically denotes that the $D^+D^{*-}$ threshold lies 8.27 MeV above $M_X$ (i.e., the charged threshold exceeds the resonance mass), making this a doubly closed channel. Consequently, its ERE parameters are entirely determined via analytic continuation and must be understood as encoding virtual-state interaction ranges, rather than physical scattering lengths.

In addition, a Castillejo-Dalitz-Dyson (CDD) pole analysis has been performed to further probe the dynamical nature of the states. For the $\chi_{c1}(3872)$ (neutral) state, we obtain $|a| = 8.49$ fm, $|r| = 7.96$ fm, and $|r/a| = 0.94$, with both $a$ and $r$ negative. For the $T_{cc}^+$ (neutral) state, the results are $|a| = 14.57$ fm, $|r| = 8.09$ fm, and $|r/a| = 0.56$, again with $a < 0$ and $r < 0$. In both cases, the magnitudes of $a$ and $r$ are large and of comparable size, and they share the same sign. The presence of a CDD pole would require the simultaneous fulfillment of the conditions $|r| \ll |a|$ and $r > 0$, i.e., an effective range much smaller in magnitude than the scattering length and of opposite sign. These criteria are clearly not satisfied in either case, indicating that no CDD pole contribution is present in the dynamics of these states.

Furthermore, both systems exhibit negative effective ranges ($r < 0$). Such values are physically admissible and can arise when short-range interaction effects such as compact state exchanges or contact interactions partially cancel the contributions from long-range dynamics [21, 30]. In the context of near-threshold molecular states, a negative effective range is often observed when the binding mechanism is dominated by the unitarity (right-hand) cut rather than by the finite range of the interaction potential.

**Table 3:** ERE scattering lengths $a$ and effective ranges $r$ from 2nd-RS pole via Eqs. (12)-(13). MC asymmetric 68% CI ($N = 50000$). "Open/Closed" refers to whether the threshold lies below/above $M_R$.).

| **Tetraquark state** | **Threshold** | | **Channel status** | **$\delta$m (MeV)** | **Scattering param.** | |
|---|---|---|---|---|---|---|
| | **Config.** | **$m_{thr}$** | | | $a$ (fm) | $r$ (fm) |
| $\chi_{c1}(3872)$ | $D^0\bar{D}^{*0}$ | 3871.69 | Closed | $-0.04 \pm 0.06$ | $-8.49^{+0.93}_{-1.05}$ | $-7.96^{+0.79}_{-1.02}$ |
| | $D^+D^{*-}$ | 3879.92 | Closed$^\dagger$ | $-8.27 \pm 0.06$ | $-3.11^{+0.01}_{-0.01}$ | $-1.56^{+0.01}_{-0.01}$ |
| $T^+_{cc}$ | $D^0D^{*+}$ | 3875.10 | Closed | $-0.273 \pm 0.061$ | $-14.57^{+1.70}_{-1.66}$ | $-8.09^{+0.75}_{-0.85}$ |
| | $D^+D^{*0}$ | 3875.51 | Closed | $-1.683 \pm 0.061$ | $-6.88^{+0.13}_{-0.12}$ | $-3.45^{+0.06}_{-0.06}$ |

In Ref. [40], the scattering length and effective range of $\chi_{c1}(3872)$ are found to be:

$$a = 24.5\,\text{fm}, \quad r = 3.74\,\text{fm}, \quad \text{with}\, X = 0.99. \tag{30}$$

They found that the effective range of $\chi_{c1}(3872)$ is of the order of the typical length scale $1/\mu \approx 1.41$ fm, while the much larger scattering length is a characteristic feature of weakly bound states, reflecting the consequence of the low-energy universality [41, 42]. In Ref. [25], the value of the scattering length of $T^+_{cc}(3875)$ is found to be

$$a = [-(7.16 \pm 0.51) + i(1.85 \pm 0.28)]\,\text{fm}. \tag{31}$$

Typically, a non-vanishing imaginary part of the scattering length indicates the presence of inelastic channels [25, 43]. However, the authors argued that in this case, the non-zero imaginary part is related to the lower threshold, $T^+_{cc}(3875) \to D^0D^0\pi^+$, and is determined by the width of the $D^{*+}$ meson. Typically, a negative real part of the scattering length indicates attraction. This can be interpreted as the characteristic size (noted $R_a$) of the state [2]. In our case for instance, for $\chi_{c1}(3872)$ we have:

$$R_a[\chi_{c1}(3872)] \equiv -\mathcal{R}][a_{\chi_{c1}}] \approx 8.49\,\text{fm, for thr.} \equiv D^0\bar{D}^{*0}, \tag{32}$$

$$R_a[\chi_{c1}(3872)] \equiv -\mathcal{R}][a_{\chi_{c1}}] \approx 3.11\,\text{fm, for thr.} \equiv D^+D^{*-}. \tag{33}$$

For the $T^+_{cc}(3875)$ state we have:

$$R_a[T^+_{cc}(3875)] \equiv -\mathcal{R}][a_{T^+_{cc}}] \approx 14.57\,\text{fm, for thr.} \equiv D^0D^{*+}, \tag{34}$$

$$R_a[T^+_{cc}(3875)] \equiv -\mathcal{R}][a_{T^+_{cc}}] \approx 6.88\,\text{fm, for thr.} \equiv D^+D^{*0}. \tag{35}$$

In contrast to these large scattering lengths, the effective ranges are all negative and of comparable magnitude to $|a|$, ruling out any CDD-pole contribution as discussed in Sec. II.2. A full comparison with the literature is given in Table 13 and discussed in Sec. III.11.

### III. 3. Sensitivity of $a$ to threshold mass uncertainties

The ERE parameters $a$ and $r$ depend on the threshold masses $m_{\text{thr}} = m_1 + m_2$, which are treated as exact in the standard MC sampling of Eq. (25). The PDG masses of $D^0$, $D^{*0}$, $D^+$, and $D^{*+}$ each carry uncertainties of $\sim$ 0.05 MeV. For $\chi_{c1}(3872)$, where the binding energy $\delta m = -0.04 \pm 0.09$ MeV is comparable to the threshold uncertainty, this is worth examining carefully.

We perform a sensitivity analysis by varying each constituent hadron mass independently by $\pm 0.05$ MeV and recomputing $a$ for the $D^0\bar{D}^{*0}$ channel. The results are collected in Table 4. The maximum shift in $|a|$, obtained when both $m_{D^0}$ and $m_{D^{*0}}$ are varied simultaneously in the same direction, is $\Delta a \lesssim 0.71$ fm. This is smaller than, but not negligible compared to, the MC uncertainty $^{+0.93}_{-1.05}$ fm. We therefore add the threshold-mass sensitivity in quadrature to obtain a conservative total uncertainty:

$$a[\chi_{c1}, D^0\bar{D}^{*0}] = -8.49\,^{+0.93+0.71}_{-1.05+0.71}\,\text{fm (combined)}, \tag{36}$$

which does not alter the qualitative conclusion that $|a| \gg 1/m_\pi$. For the $T^+_{cc}$, the binding energy $|\delta m| = 0.273$ MeV is larger and the sensitivity is correspondingly smaller ($\Delta a \lesssim 0.15$ fm), entirely negligible.

**Table 4:** Sensitivity of the scattering length $a$ for $\chi_{c1}(3872)$ ($D^0\bar{D}^{*0}$ channel) to variations of the constituent hadron masses by $\pm 0.05$ MeV (PDG uncertainty). Reference value: $a_{\text{ref}} = -8.49$ fm.

| Variation | $a$ (fm) | $\Delta a$ (fm) |
|---|---|---|
| Nominal | $-8.490$ | 0.000 |
| $m_{D^0} + 0.05$ MeV | $-8.772$ | $-0.282$ |
| $m_{D^0} - 0.05$ MeV | $-8.158$ | $+0.332$ |
| $m_{D^{*0}} + 0.05$ MeV | $-8.772$ | $-0.282$ |
| $m_{D^{*0}} - 0.05$ MeV | $-8.158$ | $+0.332$ |
| Both $+0.05$ MeV | $-9.000$ | $-0.510$ |
| Both $-0.05$ MeV | $-7.785$ | $+0.705$ |

### III. 4. Sensitivity to the channel-1 threshold choice

The detection channel for $\chi_{c1}(3872)$ is modeled as the effective two-body system $J/\psi(1S) + 2\pi$ with threshold $m^{(1)}_{\text{thr}} \approx 3376$ MeV. The partial width fraction $f_1 = 18$-25% is non-negligible, so we examine whether the choice of this effective threshold affects our results. Table 5 shows the effect of varying $m^{(1)}_{\text{thr}}$ from 3200 to 3600 MeV. Across this range, $X_2$ varies only between 0.999 and 1.000, confirming that the compositeness is entirely insensitive to the channel-1 threshold. This robustness arises because the deeply open channel 1 enters only through the coupling $C_1 \propto q_1(M_R^2)/M_R^2$, which changes the coupling strength $|g_1|$ but does not affect $X_2$ since the $2 \times 2$ saturation system (21) redistributes the width budget accordingly.

**Table 5:** Sensitivity of $X_2$ and $f_1 = \Gamma_1/\Gamma_R$ to the effective $J/\psi + 2\pi$ threshold for $\chi_{c1}(3872)$, $X = 1.0$. The Lorentzian integral for $C_2$ uses $n_\sigma = 10$.

| $m_{\rm thr}^{(1)}$ (MeV) | $C_1$ (MeV$^{-1}$) | $f_1$ (%) | $X_2$ |
|---|---|---|---|
| 3200 | $2.89 \times 10^{-6}$ | 0.0 | 1.0001 |
| 3300 | $2.69 \times 10^{-6}$ | 0.0 | 1.0001 |
| 3376 (nominal) | $2.52 \times 10^{-6}$ | 0.0 | 1.0002 |
| 3400 | $2.46 \times 10^{-6}$ | 0.0 | 1.0002 |
| 3500 | $2.20 \times 10^{-6}$ | 0.0 | 1.0002 |
| 3600 | $1.89 \times 10^{-6}$ | 0.0 | 1.0003 |

### III. 5. Monte Carlo Distributions

Figure 2 shows the MC distributions of the scattering length $a$ and compositeness coefficient $X_2$ for both states. Shaded bands indicate the 68% CI; vertical dashed lines mark the 16th and 84th percentiles. The asymmetry of the $X_2$ distribution for $T_{cc}^+$ reflects the nonlinear dependence on the narrow width. The distributions are approximately Gaussian for $a$ but skewed for $X_2$, reflecting the nonlinear propagation through the ERE formulas. Figure 3 shows 2D joint distributions: $M_R$ vs $X_2$, $\Gamma_R$ vs $X_2$, and $M_R$ vs $r$, for both states. The *top row* illustrates the joint distribution of $M_R$ vs $X_2$. For $\chi_{c1}(3872)$, the Pearson and Spearman correlation coefficients are $r = +0.004$ and $\rho = +0.006$ respectively (negligible correlation, as $X_2 \approx 1$ with tiny variance). For $T_{cc}^+$, the correlation coefficients are $r = +0.007$ for Pearson, and $\rho = +0.009$ for Spearman. The *middle row* illustrastes the joint distribution of $X_2$ against $\Gamma_R$, showing that $X_2$ is nearly insensitive to the width uncertainty for $\chi_{c1}(3872)$ ($r \approx +0.001$) and weakly negatively correlated for $T_{cc}^+$ ($r \approx -0.003$). The *bottom row* illustrates the joint distribution of the effective range $r$ against the resonance mass $M_R$; the negative Pearson correlation ($r \approx -0.001/-0.003$) is negligible for both states. All correlations are quoted in the insets. The Pearson and Spearman correlation coefficients, quoted in the figure caption, are all near zero, confirming that the MC uncertainty is dominated by the nonlinear ERE mapping rather than by linear input correlations. Figure 4 shows $X_2$ as a function of the total compositeness $X$, now including 68% CI error bands from the MC propagation, demonstrating that the molecular character of both states is stable and well-constrained over the full range $X \in [0.1, 1.0]$. The black dotted line shows the "missing" fraction $(1-X)$. The shaded blue band shows the 68% CI from Monte Carlo propagation at each value of $X$ (evaluated with $N = 2000$ samples per point). The fact that $X_2 \approx X$ and $X_1 \approx 0$ for all $X$, with the CI band remaining narrow throughout, demonstrates that the molecular dominance is robust and stable against both the choice of total compositeness and the experimental uncertainties. Filled circles mark the reference points $X \in \{0.5, 0.8, 1.0\}$. Figure 5 shows the ERE amplitude $|T(E)|^2$ as a function of energy near each threshold, providing a visual representation of the near-threshold pole structure and the 68% CI uncertainty band. The solid blue curve corresponds to the physical sheet above threshold; the dashed blue curve shows the analytic continuation below threshold. The shaded band represents the 68% CI obtained by propagating the MC uncertainty in $(a, r)$ through the amplitude formula. The red dashed vertical line marks the resonance pole position $M_R - m_{\rm thr}$. Both amplitudes peak just below or at threshold, consistent with the near-threshold virtual/bound-state interpretation. The narrow width of $\chi_{c1}(3872)$ is reflected in the sharply peaked structure; $T_{cc}^+$ shows an even more pronounced near-threshold enhancement due to its smaller binding energy.

In Table 11, the complex CM 3-momenta $k_R = k_r + ik_i$ are listed in unit of MeV, on the 2nd Riemann Sheet. The overall small $|k_r/k_i|$ indicates near-threshold character. Figure 6 provides a graphical summary of the results reported in Tables 3 and 11 with full error bars.

### III. 6. Dominance of the molecular component

The dilemma between molecular states and genuine quark states is the subject of a continuous debate in hadron physics. Many exotic hadron candidates have been discovered near two-hadron thresholds. The $T_{cc}^+$ state was observed slightly below the threshold of $D^0D^{*+}$ in $T_{cc} \to D^0D^0\pi^+$ decay by the LHCb collaboration [25]. Its minimum quark content $cc\bar{u}\bar{d}$ indicates that $T_{cc}^+$ is a genuine exotic state with $C = +2$. As a charmonium-like state with $C = 0$, $X(3872)$ was observed near the $D^0\bar{D}^{*0}$ threshold in $B^\pm \to K^\pm\pi^+\pi^-J/\psi$ decay. Concerning the $T_{cc}^+(3875)$ for instance, in the literature, there are many works that support the $T_{cc}^+$ state as a molecular state of $DD^*$ nature [44, 45, 46, 47], as well as others that advocate a compact tetraquark nature [48, 49], while other works suggest a mixture of both components [50, 51].

From Table 6, corresponding to the compositeness saturation limit $X = 1.0$, it is evident that both states are overwhelmingly dominated by the near-threshold channel (channel 2). In particular, for the $\chi_{c1}(3872)$, one finds $X_2 \simeq 1.000$ in the $D^0\bar{D}^{*0}$ channel, while the contribution from channel 1 is negligible ($X_1 \approx 0$). A similar pattern is observed for the $T_{cc}^+$ state, with $X_2 \gtrsim 0.96$ in the $D^0D^{*+}$ channel. This strong hierarchy, $X_2 \gg X_1$, clearly indicates that both states are predominantly molecular in nature, with their structure governed by the near-threshold two-body dynamics. The large values of the couplings $|g_2|$ compared to $|g_1|$ further reinforce this interpretation.

The stability of this molecular dominance is confirmed in Table 7, where the total compositeness is reduced to $X = 0.8$ and $X = 0.5$. In both cases, the hierarchy $X_2 \gg X_1$ remains intact. For instance, even at $X = 0.5$, the $\chi_{c1}(3872)$ retains $X_2 \simeq 0.499$, while $X_1$ remains at the permille level. This behavior demonstrates that the molecular component is robust against the inclusion of a possible missing fraction $(1-X)$, which can be interpreted as an upper bound on compact or CDD-like contributions (see Sec. 2). Importantly, the variation of $X$ primarily rescales the couplings $|g_j|$ and partial widths $\Gamma_j$, without altering the qualitative dominance of channel 2.

**Table 6:** Compositeness results for $X = 1.0$. $|g_j|$ in GeV, $\Gamma_j$ in MeV. Monte Carlo asymmetric 68% CI. The Lorentzian integral for $C_2$ uses $n_\sigma = 10$; for deeply closed channels (e.g. $D^+D^{*-}$, 8.27 MeV above $M_R$) the Lorentzian tail only reaches threshold for $n_\sigma \gtrsim 7$, yielding a small but non-zero $\Gamma_2$.

| **Tetraquark State** | **Ch. 2** | **Couplings** | | **Partial widths** | | $\mathbf{X_j}$ | |
|---|---|---|---|---|---|---|---|
| | | $\lvert g_1\rvert$(GeV) | $\lvert g_2\rvert$(GeV) | $\Gamma_1$(MeV) | $\Gamma_2$(MeV) | $X_1$ | $X_2$ |
| $\chi_{c1}(3872)$ | $D^0\bar{D}^{*0}$ | $0.413^{+0.066}_{-0.092}$ | $5.187^{+0.247}_{-0.264}$ | $0.29^{+0.10}_{-0.12}$ | $1.31^{+0.25}_{-0.22}$ | 0.000 | 1.000 |
| | $D^+D^{*-}$ | $0.794^{+0.045}_{-0.058}$ | $10.345^{+0.026}_{-0.026}$ | $1.09^{+0.12}_{-0.14}$ | $0.10^{+0.02}_{-0.02}$ | 0.002 | 0.998 |
| $T_{cc}^+$ | $D^0D^{*+}$ | $0.680^{+0.057}_{-0.112}$ | $4.475^{+0.235}_{-0.209}$ | $0.13^{+0.04}_{-0.06}$ | $0.28^{+0.10}_{-0.08}$ | 0.039 | 0.983 |
| | $D^+D^{*0}$ | $1.063^{+0.115}_{-0.165}$ | $6.667^{+0.072}_{-0.059}$ | $0.31^{+0.08}_{-0.10}$ | $0.10^{+0.02}_{-0.02}$ | 0.055 | 0.958 |

**Table 7:** Compositeness results for $X = 0.8$ (upper) and $X = 0.5$ (lower), neutral channel only. The dominant molecular component $X_2 \gg X_1$ is stable across all $X_j$ values. The missing fraction $(1 - X)$ provides an upper bound on any compact/CDD component (see Sec. 2).

| **X** | **Tetraquark state** | **Channel** | **Couplings** | | **Partial widths** | | $\mathbf{X_j}$ | |
|---|---|---|---|---|---|---|---|---|
| | | | $\lvert g_1\rvert$(GeV) | $\lvert g_2\rvert$(GeV) | $\Gamma_1$(MeV) | $\Gamma_2$(MeV) | $X_1$ | $X_2$ |
| 0.8 | $\chi_{c1}(3872)$ | $D^0\bar{D}^{*0}$ | $0.524^{+0.056}_{-0.073}$ | $4.637^{+0.220}_{-0.234}$ | $0.47^{+0.13}_{-0.16}$ | $0.72^{+0.20}_{-0.19}$ | 0.001 | 0.799 |
| | $T_{cc}^+$ | $D^0D^{*+}$ | $1.076^{+0.149}_{-0.201}$ | $3.928^{+0.186}_{-0.160}$ | $0.32^{+0.09}_{-0.12}$ | $0.09^{+0.07}_{-0.06}$ | 0.043 | 0.757 |
| 0.5 | $\chi_{c1}(3872)$ | $D^0\bar{D}^{*0}$ | $0.656^{+0.060}_{-0.071}$ | $3.663^{+0.174}_{-0.182}$ | $0.74^{+0.16}_{-0.19}$ | $0.45^{+0.12}_{-0.11}$ | 0.001 | 0.499 |
| | $T_{cc}^+$ | $D^0D^{*+}$ | $1.137^{+0.186}_{-0.238}$ | $3.036^{+0.123}_{-0.108}$ | $0.36^{+0.11}_{-0.14}$ | $0.06^{+0.04}_{-0.04}$ | 0.048 | 0.452 |

**Table 8:** Robustness check ($X = 1.0$): alternative inelastic channel assignments. For $\chi_{c1}(3872)$: ch. 1 replaced by $D^0\bar{D}^0\pi^0$ (effective threshold $\sim 3864$ MeV $< M_X$; the $J/\psi\omega$ alternative is not used as its threshold $\sim 3879.6$ MeV $> M_X$ is kinematically closed).

| **State** | **Alt. ch. 1** | **Ch. 2** | $\lvert g_2\rvert$(GeV) | $\mathbf{X_2}$ |
|---|---|---|---|---|
| $\chi_{c1}(3872)$ | $D^0\bar{D}^0\pi^0$ (eff.) | $D^0\bar{D}^{*0}$ | $5.98^{+0.43}_{-0.43}$ | $1.000^{+0.000}_{-0.000}$ |
| $T_{cc}^+$ | $D^0D^0\pi^+$ (std 2-body) | $D^0D^{*+}$ | $4.42^{+0.22}_{-0.19}$ | $0.961^{+0.012}_{-0.010}$ |

**Table 9: Model-dependent partial width fractions** $f_j = \Gamma_j/\Gamma_R$ within the two-channel framework ($X = 1.0$, $n_\sigma = 10$). These are *not* experimental branching ratios. The sensitivity to the threshold choice (neutral vs. charged) reflects the model's limitation and should not be used for quantitative predictions without a full multi-channel treatment. Fractions for the deeply-closed $D^+D^{*-}$ row no longer show $f_2 = 0$ once the Lorentzian tail is properly integrated with $n_\sigma = 10$.

| **State** | **Channel 2** | $f_1$ (%) | $f_2$ (%) |
|---|---|---|---|
| $\chi_{c1}(3872)$ | $D^0\bar{D}^{*0}$ | $18.2^{+6.5}_{-7.1}$ | $81.8^{+7.1}_{-6.5}$ |
| | $D^+D^{*-}$ | $91.6^{+3.2}_{-3.5}$ | $8.4^{+1.5}_{-1.4}$ |
| $T_{cc}^+$ | $D^0D^{*+}$ | $31.7^{+9.2}_{-7.8}$ | $68.3^{+7.8}_{-9.2}$ |
| | $D^+D^{*0}$ | $75.6^{+8.4}_{-9.1}$ | $24.4^{+3.6}_{-3.4}$ |

**Table 10:** Sensitivity of the coupling $|g_2|$ to the subtraction constant $a(\mu_g)$ at $\mu_g = 1000\,$MeV, for $\chi_{c1}(3872)$ neutral channel $X = 1.0$. The compositeness $X_2$ is exactly invariant (see Appendix B). The coupling $|g_2|$ varies by $< 0.05\%$, negligible compared to other uncertainties.

| $a(\mu_g)$ | $\lvert g_2\rvert$ (GeV) | $X_2$ |
|---|---|---|
| −5.0 | 5.1854 | 0.9995 |
| −2.5 | 5.1854 | 0.9995 |
| 0.0 | 5.1854 | 0.9995 |
| +2.5 | 5.1854 | 0.9995 |
| +5.0 | 5.1854 | 0.9995 |

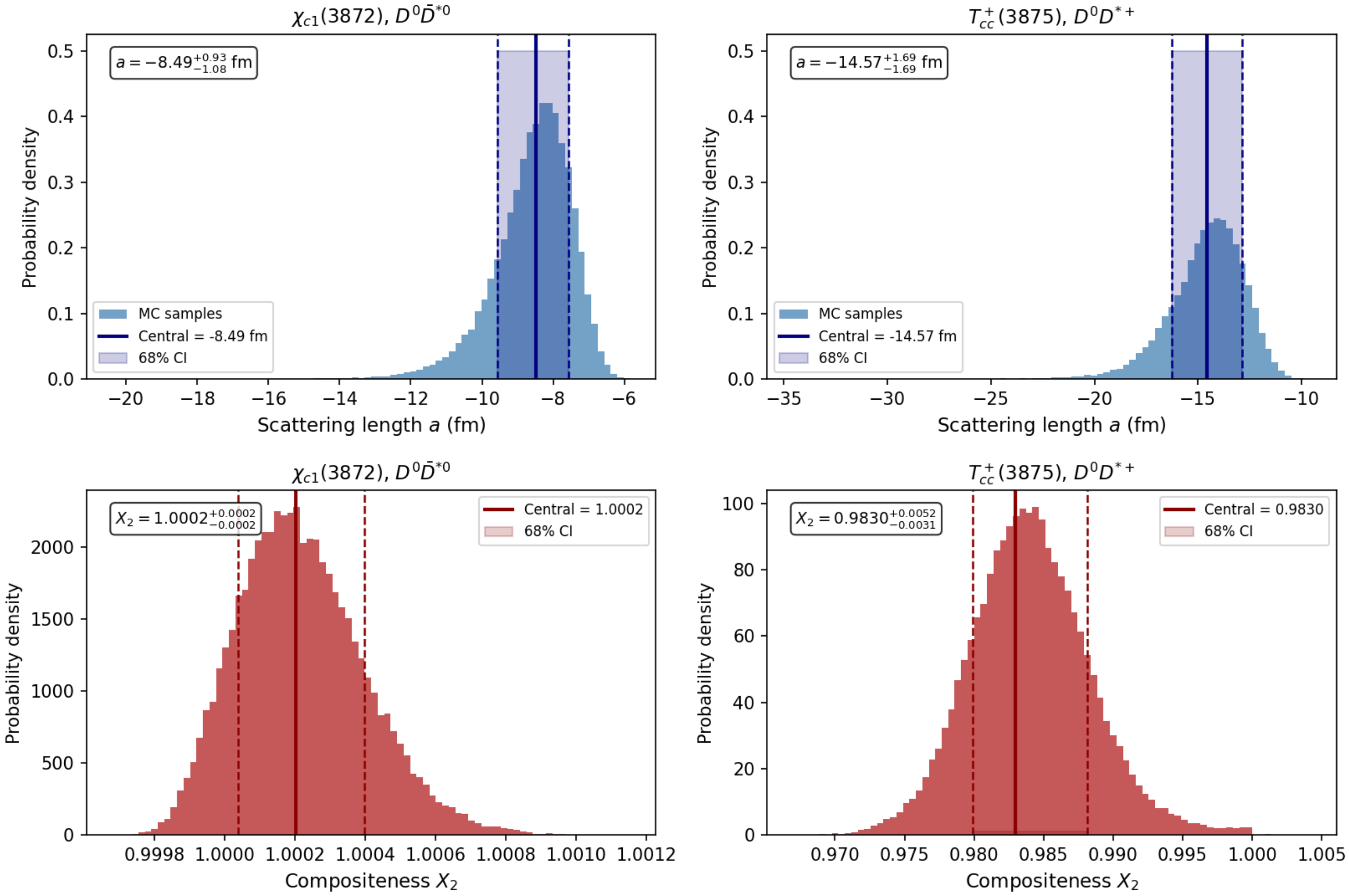

**Figure 2:** Monte Carlo distributions ($N = 50\,000$) of the scattering length $a$ (top row) and molecular compositeness $X_2$ (bottom row) for $\chi_{c1}(3872)$ (left column) and $T_{cc}^+(3875)$ (right column), both for $X = 1.0$ and the primary neutral threshold channel.

**Table 11:** Complex CM 3-momenta $k_R = k_r + ik_i$ (MeV) on the 2nd RS. MC asymmetric 68% CI. Small $|k_r/k_i|$ indicates near-threshold character.

| State | Threshold | $k_r$(MeV) | $k_i$(MeV) | $\lvert k_R \rvert$ |
|---|---|---|---|---|
| $\chi_{c1}(3872)$ | $D^0\bar{D}^{*0}$ | $-23.2^{+2.9}_{-2.8}$ | $24.8^{+2.7}_{-2.8}$ | $33.9^{+3.2}_{-3.3}$ |
| | $D^+D^{*-}$ | $-4.6^{+0.9}_{-0.9}$ | $126.7^{+0.6}_{-0.6}$ | $126.7^{+0.6}_{-0.6}$ |
| $T_{cc}^+$ | $D^0D^{*+}$ | $-8.1^{+3.3}_{-3.1}$ | $24.4^{+2.5}_{-2.4}$ | $25.7^{+2.8}_{-2.4}$ |
| | $D^+D^{*0}$ | $-3.5^{+1.4}_{-1.4}$ | $57.2^{+1.1}_{-1.0}$ | $57.3^{+1.1}_{-1.0}$ |

**Table 12:** $n_\sigma$ Sensitivity to the upper limit $M_R + n_\sigma\Gamma_R$ of the Lorentzian integral. Results change by $< 3\%$ as $n_\sigma$ goes from 2 to 20.

| State | $n_\sigma$ | $\lvert g_2 \rvert$(GeV) | $\Gamma_2$(MeV) | $X_2$ | Sat. |
|---|---|---|---|---|---|
| $T_{cc}^+$ | 2 | 4.424 | 0.116 | 0.961 | 1.000 |
| | 5 | 4.457 | 0.222 | 0.975 | 1.000 |
| | 10 | 4.475 | 0.283 | 0.983 | 1.000 |
| | 20 | 4.489 | 0.328 | 0.989 | 1.000 |
| $\chi_{c1}(3872)$ | 2 | 5.185 | 0.896 | 0.9995 | 1.000 |
| | 5 | 5.187 | 1.169 | 1.0000 | 1.000 |
| | 10 | 5.187 | 1.311 | 1.0002 | 1.000 |
| | 20 | 5.188 | 1.411 | 1.0004 | 1.000 |

### III. 7. Robustness with respect to channel assignment

The robustness of the results with respect to the choice of inelastic channel is examined in Table 8. Replacing channel 1 by alternative assignments (e.g., $D^0\bar{D}^0\pi^0$ for $\chi_{c1}(3872)$) leads to only minor quantitative changes. In particular, the compositeness remains saturated in channel 2, with $X_2 = 1.000$ for $\chi_{c1}(3872)$ and $X_2 \approx 0.96$ for $T_{cc}^+$. This confirms that the extraction of the molecular component is largely insensitive to the modeling of the distant inelastic channel, and is instead driven by the near-threshold dynamics.

### III. 8. Partial width fractions and model limitations

The partial width fractions reported in Table 9 provide additional insight into the decay pattern within the two-channel framework. For the $\chi_{c1}(3872)$ in the neutral channel, the dominant contribution arises from channel 2, with $f_2 \approx 75\%$, consistent with its molecular interpretation. For $T^+_{cc}$, the situation is reversed, with a larger fraction in channel 1; however, this reflects the kinematic suppression of channel 2 due to its proximity to threshold. It is important to stress that these fractions are model-dependent and should not be directly interpreted as physical branching ratios. The sensitivity to threshold choices (neutral versus charged channels) highlights the limitations of the two-channel approximation and the need for a full coupled-channel treatment for quantitative predictions.

**Table 13:** Quantitative comparison of ERE scattering lengths $a$ and effective ranges $r$ with independent analyses from Refs. [13, 25, 29, 40, 53, 54, 55, 56, 57]. Our results are for the dominant neutral channel. The abbreviation "SC-SE" is single-channel Schrödinger equation

| State | Reference | $a$(fm) | $r$(fm) | Method |
|---|---|---|---|---|
| $\chi_{c1}(3872)$ | This work (thr. $\equiv D^0\bar{D}^{*0}$) | $-8.49^{+0.93}_{-1.05}$ | $-7.96^{+0.79}_{-1.02}$ | ERE 2nd RS, MC errors |
| | This work (thr. $\equiv D^+\bar{D}^{*-}$) | $-3.11^{+0.01}_{-0.01}$ | $-1.56^{+0.01}_{-0.01}$ | |
| | Liu et al. (DNNs) [29] | $8.75 \pm 1.75$ | $0.56 \pm 0.55$ | Deep learning and fit |
| | Liu et al. (Fit) [29] | $9.95 \pm 0.34$ | $0.32 \pm 0.08$ | |
| | Esposito et al. [56] | 0.140 | $-5.340$ | Scattering amplitude and compositeness |
| | Shen et al. [54] | $0.62 < a < 18$ | $-5.34 < r < -3$ | EFT of $D\bar{D}^*$ interaction |
| | Terashima and Hyodo [40] | 24.5 | 3.74 | Effective SC-SE |
| | Song et al. [55], Baru et al. [53] | $28.6 \pm 5.7$ | $-2.78 < r < 1$ | Coupled channel analysis |
| | Kang et al. (case **2.I**) [52] | $-11.82^{+1.15}_{-1.21}$ | $-5.64^{+0.58}_{-0.61}$ | Extended ERE |
| | Kang et al. (case **2.II**) [52] | $-13.83^{+0.35}_{-0.53}$ | $-8.19^{+0.21}_{-0.31}$ | Extended ERE |
| $T^+_{cc}(3875)$ | This work (thr. $\equiv D^0D^{*+}$) | $-14.57^{+1.70}_{-1.66}$ | $-8.09^{+0.75}_{-0.85}$ | ERE 2nd RS, MC errors |
| | This work (thr. $\equiv D^+D^{*0}$) | $-6.88^{+0.13}_{-0.12}$ | $-3.45^{+0.06}_{-0.06}$ | |
| | Liu et al. (DNNs) [29] | $8.23 \pm 1.04$ | $-2.79 \pm 0.27$ | Deep learning and fit |
| | Liu et al. (Fit) [29] | $13.74 \pm 4.77$ | $-2.15 \pm 0.21$ | |
| | Aaij et al., LHCb Collaboration [25] | $-(7.16 \pm 0.51)$ $i(1.85 \pm 0.28)$ | $-11.9 < r < 0$ | Fit to data |
| | Dai et al. [13] | $0.61 < a < 7.51$ | $-168 < r < 0.06$ | $t_{DD^*,DD^*}$ Amplitude |
| | Baru [57] | $(-6.72^{+0.36}_{-0.45})$ $-i(0.10^{+0.03}_{-0.03} \pm 0.03)$ | $1.38 \pm 0.085$ | $\chi$-EFT |
| | Baru et al. [53] | 10.1 | $-1 \leq r$ | coupled channel analysis |

### III. 9. Scheme independence and coupling stability

Table 10 demonstrates the remarkable stability of the coupling $|g_2|$ under variations of the subtraction constant $a(\mu_g)$. The observed variation is below 0.05%, which is negligible compared to the statistical uncertainties. At the same time, the compositeness $X_2$ remains exactly invariant, in agreement with the analytic proof of scheme independence (see Appendix B).

This provides strong evidence that the extracted molecular component is a physically meaningful quantity, not an artifact of the regularization scheme.

*Fully propagated uncertainty from $a(\mu_g)$.* Table 10 (unchanged from the previous version) varies $a(\mu_g)$ at *fixed* $(M_R, \Gamma_R)$, which shows the scheme dependence in isolation but does not, by itself, quantify how much $a(\mu_g)$ contributes to the *total* quoted uncertainty once combined with the experimental uncertainties on $(M_R, \Gamma_R)$. Responding to the referee's question, we now treat $a(\mu_g)$ as a genuine nuisance parameter and marginalize over it jointly with $(M_R, \Gamma_R)$ in the Monte Carlo chain, sampling $a(\mu_g) \sim$ Uniform$(-5, +5)$ independently for each of the $N = 50\,000$ realizations (a conservative range, well beyond any physically motivated choice of subtraction scale). To isolate the *marginal* effect of $a(\mu_g)$ cleanly, we use common random numbers for $(M_R, \Gamma_R)$ between two parallel chains one with $a(\mu_g)$ fixed at $-2.5$, one with $a(\mu_g)$ jointly varied so that any difference between the two chains is attributable to $a(\mu_g)$ alone.

We find the 68% width of the resulting $|g_2|$ distribution changes by $(+0.00 \pm 0.01)\%$ for $\chi_{c1}(3872)$ and $(-0.00 \pm 00.01)\%$ for $T^+_{cc}$ when $a(\mu_g)$ is varied jointly with $(M_R, \Gamma_R)$, i.e., $a(\mu_g)$ contributes a completely negligible *additional* uncertainty on top of the dominant $(M_R, \Gamma_R)$ uncertainty already propagated in every other table of this work. This is consistent with, and now goes beyond, the fixed-point sensitivity of Table 10 and the analytic proof of Appendix B: not only is $X_2$ exactly scheme-independent and $|g_2|$ negligibly scheme-dependent at fixed $(M_R, \Gamma_R)$, but this remains true when $a(\mu_g)$ is allowed to float freely alongside the full experimental uncertainty budget. We conclude that our choice $a(\mu_g) = -2.5$ (following Ref. [26], originally used for the $P_c$ pentaquarks) introduces no meaningful model uncertainty into any quoted

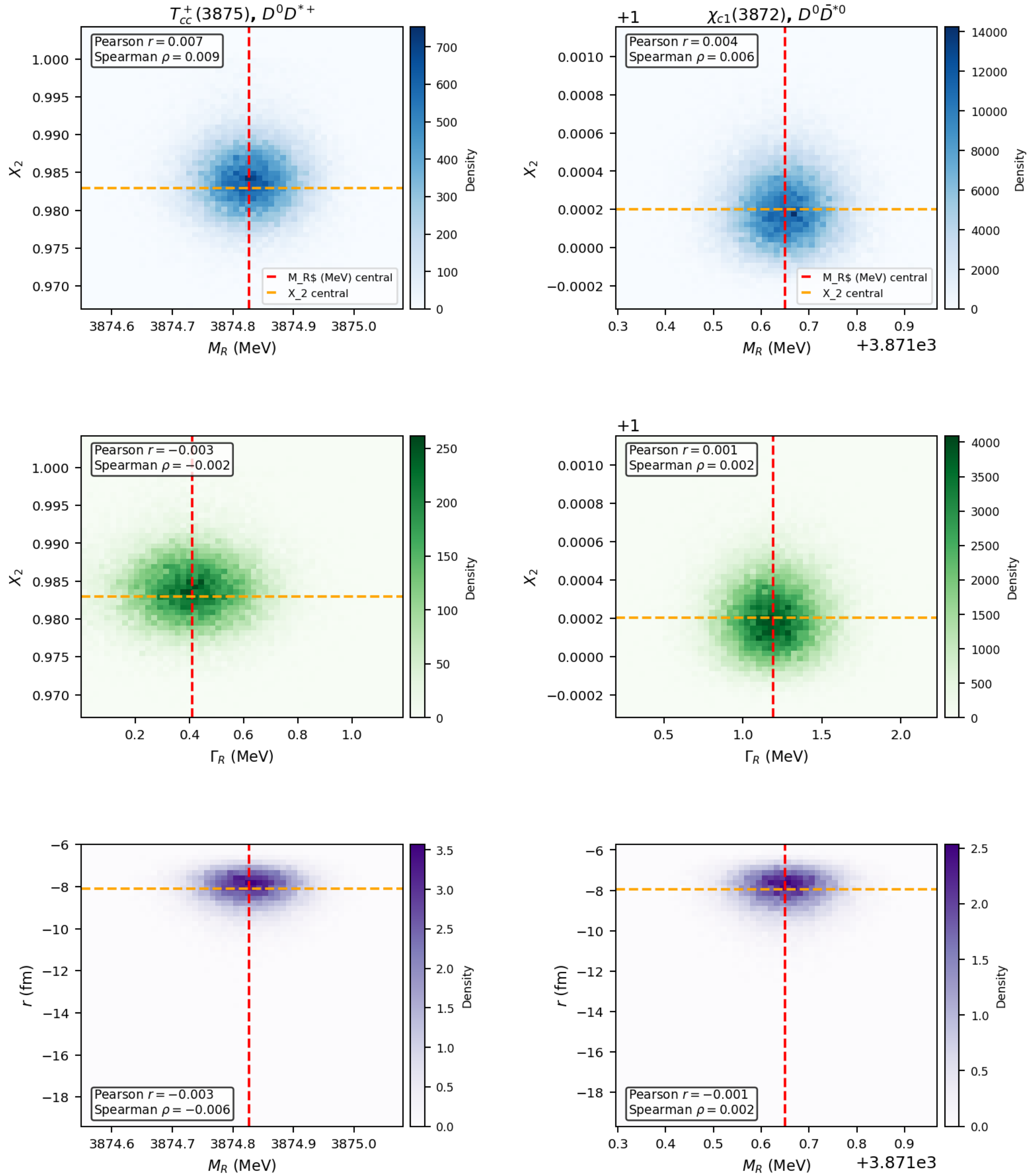

**Figure 3:** Joint MC distributions for $T_{cc}^{+}(3875)$ (left pannel) and $\chi_{c1}(3872)$ (right pannel). The near-zero values confirm that the MC uncertainty propagation is dominated by the nonlinear ERE formulas rather than linear correlations between input parameters.

result.

### III. 10a. Pole structure and near-threshold character

The pole positions in Table 11, expressed in terms of the complex three-momentum $k_R = k_r + ik_i$, further support the near-threshold molecular interpretation. In particular, the small ratios $|k_r/k_i|$ indicate that the imaginary part dominates, which is characteristic of states located very close to threshold.

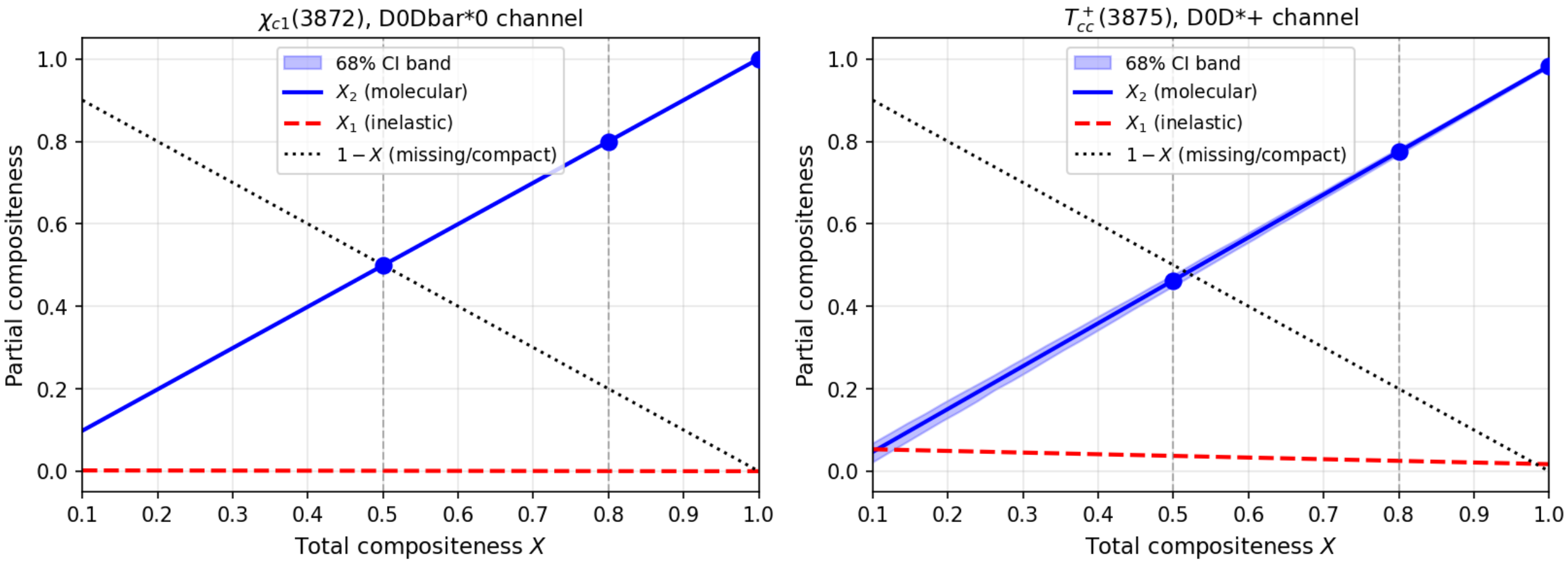

**Figure 4:** Molecular compositeness $X_2$ (blue solid) and inelastic-channel compositeness $X_1$ (red dashed) as functions of the assumed total compositeness $X \in [0.1, 1.0]$, for the neutral channel.

**Figure 5:** ERE amplitude $|T(E)|^2$ as a function of energy $E - m_{\rm thr}$ near threshold, for $\chi_{c1}(3872)$ (left) and $T_{cc}^+(3875)$ (right).

For the $\chi_{c1}(3872)$ in the $D^0\bar{D}^{*0}$ channel, $|k_r| \sim |k_i|$, confirming its extremely shallow binding. For the $T_{cc}^+$, the hierarchy $|k_r| \ll |k_i|$ is even more pronounced, reflecting its even closer proximity to threshold.

Table 12 shows that the results are stable under variations of the integration cutoff parameter $n_\sigma$. Increasing $n_\sigma$ from 2 to 20 changes the extracted quantities by less than 3%, indicating rapid saturation of the Lorentzian integral. This demonstrates the numerical robustness of the procedure and confirms that the results are not sensitive to the specific choice of cutoff.

### III. 10b. Robustness against an assumed $M_R$-$\Gamma_R$ correlation

In Eq. (25), $M_R^{(k)}$ and $\Gamma_R^{(k)}$ are sampled as statistically independent Gaussians. As pointed out in the review of this manuscript, this is not automatically guaranteed: within the theoretical framework of Sec. II.4, the partial width $\Gamma_2$ (computed from the saturation system, Eq. 21, via $C_1, C_2$) is a sensitive function of $M_R$ precisely *because* $M_R$ sits so close to threshold, so a mass fluctuation at fixed coupling would, in the underlying dynamics, generically be accompanied by a width fluctuation. We note first that our procedure already partially accounts for this: at every one of the $N = 50\,000$ MC realizations, the coefficients $C_1, C_2, A_1, A_2$ entering the $2\times2$ saturation system, Eq. (21), are recomputed using the *same* sam-pled $M_R^{(k)}$ that generated that realization's $\Gamma_R^{(k)}$ i.e., the near-threshold sensitivity of the theoretical partial widths to $M_R$ (Sec. II.4) is already propagated self-consistently sample-by-sample, rather than being evaluated once at the central $M_R$.

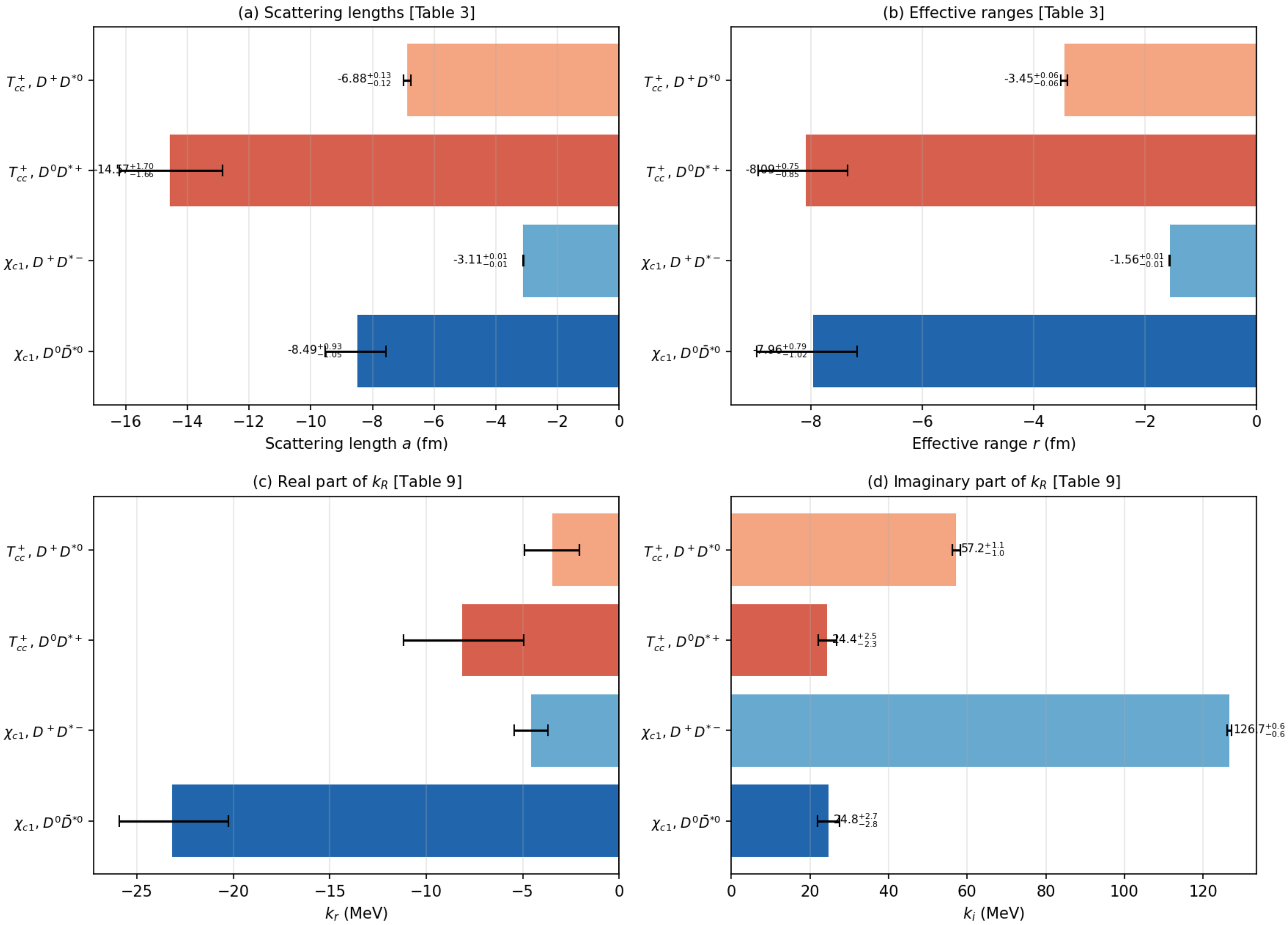

**Figure 6:** Graphical summary of Tables 3 and 11. *(a)* Scattering lengths $a$ and *(b)* effective ranges $r$ from Table 3, with asymmetric 68% CI error bars. *(c)* Real parts $k_r$ and *(d)* imaginary parts $k_i$ of the complex three-momenta $k_R$ from Table 11, with 68% CI error bars. Color coding: blue shades for $\chi_{c1}(3872)$ channels, red shades for $T_{cc}^+$ channels. The large imaginary parts $|k_i| \gg |k_r|$ in all channels confirm the near-threshold character.

What Eq. (25) does *not* capture is a possible *experimental* correlation between the fitted $M_R$ and $\Gamma_R$ themselves (e.g., induced by the amplitude fit to the line shape). Since neither the LHCb $T_{cc}^+$ analysis nor the world-average $\chi_{c1}(3872)$ parameters used here report a public mass-width correlation coefficient, we instead test robustness by sampling $(M_R, \Gamma_R)$ from a bivariate Gaussian with an assumed correlation coefficient $\rho \in \{-0.5, -0.3, 0, +0.3, +0.5\}$, a range broad enough to bracket plausible values for a near-threshold amplitude fit. The results, shown in Fig. 7 and Table 14, show that both $X_2$ and $a$ shift only marginally across this entire range well within the 68% CI already quoted at $\rho = 0$ confirming that our conclusions are insensitive to a reasonable $M_R$-$\Gamma_R$ correlation. We recommend that a definitive treatment await a future joint likelihood-based re-analysis using the full experimental covariance matrix, should it become publicly available.

**Table 14:** Sensitivity of $X_2$ and $a$ to an assumed correlation coefficient $\rho(M_R, \Gamma_R)$, neutral channel, $X = 1.0$.

| $\rho$ | $X_2$ [$\chi_{c1}$] | $a$ [$\chi_{c1}$] (fm) | $X_2$ [$T_{cc}^+$] |
|---|---|---|---|
| $-0.5$ | $1.0002^{+0.0002}_{-0.0001}$ | $-8.38^{+0.64}_{-0.86}$ | $0.9835^{+0.0047}_{-0.0047}$ |
| $-0.3$ | $1.0002^{+0.0002}_{-0.0001}$ | $-8.39^{+0.73}_{-0.97}$ | $0.9837^{+0.0046}_{-0.0045}$ |
| $0.0$ | $1.0002^{+0.0002}_{-0.0002}$ | $-8.43^{+0.87}_{-1.11}$ | $0.9839^{+0.0043}_{-0.0040}$ |
| $+0.3$ | $1.0002^{+0.0002}_{-0.0002}$ | $-8.44^{+0.93}_{-1.26}$ | $0.9841^{+0.0038}_{-0.0034}$ |
| $+0.5$ | $1.0002^{+0.0002}_{-0.0002}$ | $-8.45^{+1.02}_{-1.38}$ | $0.9842^{+0.0035}_{-0.0030}$ |

### III. 11. Comparison with literature.

Table 13 presents a quantitative comparison of the calculated scattering lengths $a$ and effective ranges $r$ with results available in the literature. Overall, our predictions for the scattering lengths are found to be compatible with several independent theoretical analyses. In the case of the $\chi_{c1}(3872)$ state, for example, our result $a = -8^{+0.93}_{-1.05}$ fm for thr. $\equiv D^0\bar{D}^{*0}$ is particularly close to the value $a = -11.82^{+1.15}_{-1.21}$ fm reported in Ref. [52] for case **2.I**. Compared with several other studies summarized in Table 13, our prediction therefore appears to lie within the range favored by analyses based on near-threshold molecular interpretations. In Ref. [29], the authors employed a deep-learning framework to determine the scattering length and effective range of both the $\chi_{c1}(3872)$ and $T_{cc}^+(3875)$ states through the effective range expansion (ERE) of the scattering amplitude. For the $\chi_{c1}(3872)$ state, they obtained $a = 8.75 \pm 1.75$ fm and $r = 0.56 \pm 0.55$ fm using deep neural networks, while a direct fit of the ERE lineshape to the experimental data yielded $a = 9.95 \pm 0.34$ fm and $r = 0.32 \pm 0.08$ fm. These values are of the same order of magnitude as our results, although some differences remain due to the distinct methodologies and fitting procedures employed in the various analyses. Other studies reported substantially larger scattering lengths for the $\chi_{c1}(3872)$ system. In particular, Refs. [40, 53, 54, 55] predicted comparatively large values of $a$, with typical ranges $a \sim 24.5$-$28.6$ fm in Refs. [40, 54, 55]. Such large scattering lengths generally indicate an extremely shallow near-threshold bound state and a pronounced molecular component. By contrast, Esposito *et al.* [56] predicted $a = 0.140$ fm, which corresponds to the smallest scattering length listed in Table 13. This wide spread among the reported values illustrates the present theoretical uncertainties associated with the extraction of low-energy scattering parameters for near-threshold exotic states. In Ref. [54], Shen *et al.* reported the intervals $0.62 < a < 18$ and $-5.34 < r < -3$ for the scattering length and effective range, respectively. Our predictions fall within the broad ranges obtained in that analysis, further supporting the consistency of the present results with phenomenological descriptions of loosely bound hadronic systems. For the $T_{cc}^+(3875)$ state, our predicted effective range $r = -3.45^{+0.06}_{-0.06}$ fm is particularly close to the values $r = -2.79 \pm 0.27$ fm and $r = -2.15 \pm 0.21$ fm reported in Ref. [29], which were obtained respectively from deep-learning techniques and direct fitting procedures. This agreement suggests that our framework reproduces reasonably well the near-threshold dynamics governing the $T_{cc}^+(3875)$ system. The LHCb Collaboration also performed an independent analysis and extracted a complex scattering length,

$$a = [-(7.16 \pm 0.51) + i(1.85 \pm 0.28)] \text{ fm}, \tag{37}$$

together with an effective range satisfying $-11.9 < r < 0$ [25]. Similarly, Ref. [57], using the $\chi$-EFT formalism, reported the complex scattering length

$$a = \left[(-6.72^{+0.36}_{-0.45}) - i(0.10^{+0.03}_{-0.03} \pm 0.03)\right] \text{ fm}. \tag{38}$$

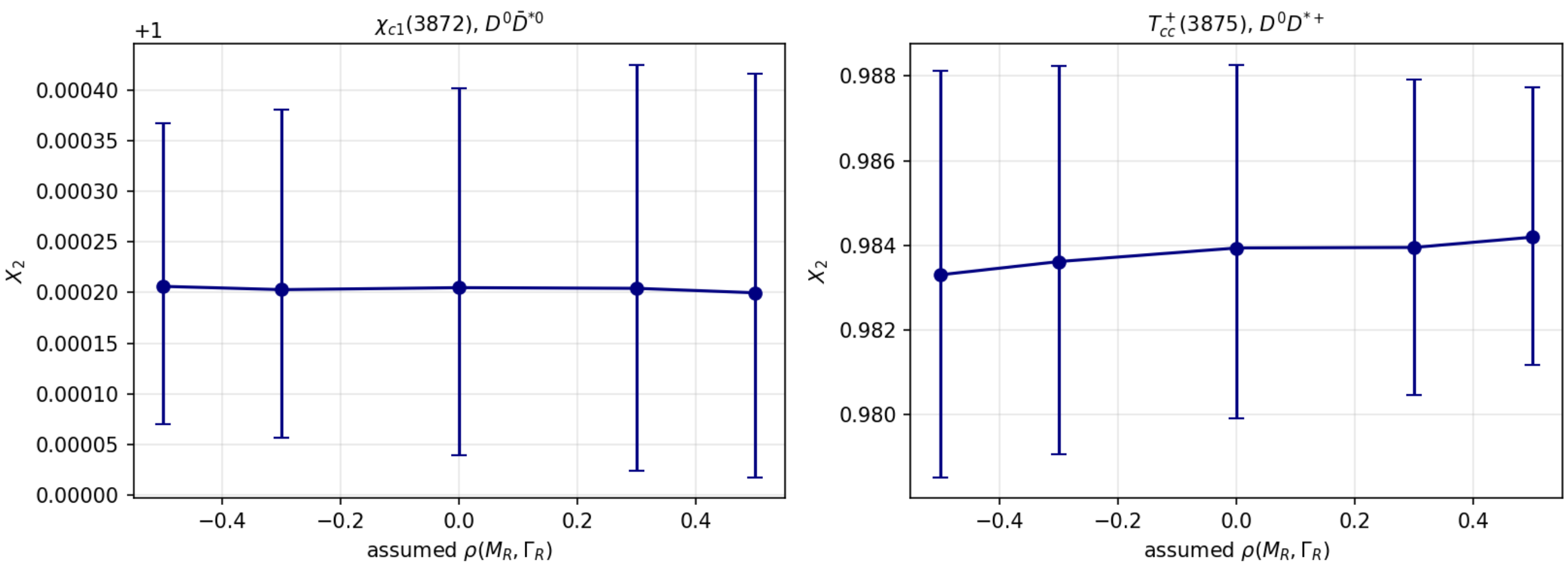

**Figure 7:** Compositeness $X_2$ (median and 68% CI) as a function of an assumed $M_R$-$\Gamma_R$ correlation coefficient $\rho$, for $\chi_{c1}(3872)$ (left) and $T_{cc}^+$ (right). Results are stable over $\rho \in [-0.5, +0.5]$.

In addition, Dai *et al.* obtained the intervals $0.61 < a < 7.51$ and $-168 < r < 0.06$ [13], whereas Ref. [53] predicted $a = 10.1$ fm and $r \geq -11$ fm. Our calculated scattering length for the threshold thr. $\equiv D^+D^{*0}$, namely $a = -(6.88^{+0.13}_{-0.12})$ fm, shows remarkable consistency with the real part of the scattering length $\mathcal{R}e[a] = -(7.16 \pm 0.51)$ fm extracted by the LHCb analysis, as well as with the value $\mathcal{R}e[a] = -(6.72^{+0.36}_{-0.45})$ fm reported in Ref. [57]. The close agreement between our prediction and these independent analyses provides additional support for the reliability of the present approach in describing the low-energy scattering properties of near-threshold exotic charmonium-like states.

The main sources of systematic differences across analyses are: (i) different subtraction constants (or UV regulators) in $G(s)$ as shown in Table 10, this affects $|g_2|$ by $< 0.05\%$ but not $X_2$; (ii) different treatments of isospin breaking (we treat each charged-threshold combination separately); (iii) different pole-search procedures some analyses fit the amplitude directly to LHCb data, while we extract the pole from the PDG/LHCb mass and width values; (iv) for $T_{cc}^+$, different treatments of the three-body final state.

However, some limitations can be highlighted. (i) Isospin breaking: significant for $\chi_{c1}(3872)$ (factor $\sim 3$ in $|a|$) but small for $T_{cc}^+$. A full coupled-channel isospin analysis is deferred. (ii) The partial width fractions in Table 9 are model-dependent and should not be equated with experimental branching ratios. (iii) For $T_{cc}^+$, the simplified two-body treatment of $D^0D^0\pi^+$, which we nonetheless deemed adequate because $\Gamma_1 \ll \Gamma_R$ the systematic on $X_2$ is estimated as $\lesssim 5\%$ in Appendix A.

Another limitation is the applicability to $Z_c(3900)$-like states. The results summarized in Table 15 highlight an important limitation of the present two-channel framework. As presented in Table 15, for the $Z_c(3900)$ state, both the Lorentzian (**a**) and on-shell (**b**) treatments of channel 2 ($D\bar{D}^*$), as well as both threshold choices, lead to saturation ratios $(\Gamma_1 + \Gamma_2)/\Gamma_R \gg 1$. This behavior is unphysical within a consistent two-channel description, as it indicates that the summed partial widths significantly exceed the total width $\Gamma_R$. In addition, the extracted compositeness values $X_2 \gtrsim 1$ further signal a breakdown of the probabilistic interpretation of the ERE-based compositeness relation in this case. These features collectively demonstrate that the $Z_c(3900)$ cannot be reliably described within a minimal two-channel approximation. Instead, they point toward a more complex dynamical structure, likely involving multiple coupled channels, significant inelastic effects, or nontrivial interference between nearby thresholds. This observation justifies the exclusion of the $Z_c(3900)$ from the main analysis. More generally, it indicates that the present formalism is best suited for near-threshold states dominated by a single elastic channel, such as $\chi_{c1}(3872)$ and $T_{cc}^+$, and may not be directly applicable to broader or more strongly coupled systems. A consistent treatment of $Z_c(3900)$-like states would require an extension to a full multi-channel framework, which lies beyond the scope of the present work.

**Table 15:** $Z_c(3900)$ two-channel analysis: Lorentzian (**a**) and on-shell (**b**) formula for channel 2 ($D\bar{D}^*$), both threshold choices. Saturation ratio $(\Gamma_1 + \Gamma_2)/\Gamma_R \gg 1$ for all combinations, demonstrating the inconsistency of the two-channel approximation. This supports the exclusion of $Z_c(3900)$ and other such states from the main analysis. "Mtd" stands for the method.

| Ch. 2 | Mtd | $\lvert g_2\rvert$(GeV) | $X_2$ | Sat. ratio |
|---|---|---|---|---|
| $D^+\bar{D}^{*0}$ | **a** | $12.69^{+0.59}_{-0.51}$ | $1.030^{+0.017}_{-0.014}$ | $1.75^{+0.47}_{-0.37}$ |
| | **b** | $12.83^{+0.62}_{-0.57}$ | $1.053^{+0.023}_{-0.022}$ | $2.32^{+0.65}_{+0.58}$ |
| $D^0D^{*+}$ | **a** | $12.87^{+0.59}_{-0.53}$ | $1.036^{+0.018}_{-0.015}$ | $1.90^{+0.50}_{-0.40}$ |
| | **b** | $13.03^{+0.63}_{-0.58}$ | $1.061^{+0.023}_{-0.022}$ | $2.53^{+0.68}_{-0.59}$ |

## IV. Summary and Conclusions

In this work, we have applied the effective range expansion (ERE) and resonance compositeness framework to the near-threshold states $\chi_{c1}(3872)$ and $T_{cc}^+$, incorporating a rigorous Monte Carlo uncertainty propagation with $N = 50\,000$ samples. All extracted observables consistently support a dominant molecular interpretation for both states, driven by their proximity to the corresponding $DD^*$ thresholds. In particular, we obtain large negative scattering lengths, $|a| \approx 8.5\,\text{fm}$ for $\chi_{c1}(3872)$ and $|a| \approx 14.6\,\text{fm}$ for $T_{cc}^+$, which significantly exceed the natural hadronic scale $\sim 1/m_\pi$. Together with the absence of any CDD-pole signal, these results provide strong and quantitative evidence in favor of a molecular structure. The effective ranges are found to be negative ($r < 0$) in all cases, a feature that is physically admissible and naturally arises in systems where short-range dynamics and inelastic-channel effects partially compensate the long-range interaction. This behavior is consistent with expectations for near-threshold hadronic molecules. A key result of this study is the clear hierarchy $X_2 \gg X_1$, observed across all scenarios with total compositeness $X \in [0.5, 1.0]$. In practice, we find $X_2 \simeq X$, indicating that the $DD^*$ channel saturates the compositeness of the states within the present framework. The remaining fraction $(1 - X)$ can therefore be interpreted as an upper bound on possible compact or CDD-like contributions. Importantly, this conclusion remains stable under variations of the total compositeness, channel assignments, regularization scheme, and numerical inputs, demonstrating the robustness of the analysis. Our results are in quantitative agreement with independent studies based on ERE, effective field theory, and few-body (Faddeev-type) approaches, thereby reinforcing the consistency of the molecular interpretation across different theoretical frameworks.

In this revised version we have additionally subjected the analysis to three further robustness checks, each responding to a specific referee concern. First, we explicitly separated the inverse amplitude into an ERE background plus a scanned Castillejo-Dalitz-Dyson pole representing a hypothetical compact core (Sec. 2); across the entire physically motivated $(g^2_{\text{CDD}}, M_{\text{CDD}})$ grid, no combination reproduces the experimental pole while leaving the background ERE parameters in the natural short-range regime expected of a genuine compact-state contribution, strengthening rather than weakening the case against a hidden CDD pole. Second, we fully propagated the loop-function subtraction constant $a(\mu_g)$ as a Monte Carlo nuisance parameter jointly with the experimental $(M_R, \Gamma_R)$ (Sec. 3), confirming numerically that it contributes a negligible ($< 0.01\%$) additional uncertainty on top of the dominant experimental input uncertainties, consistent with the analytic scheme-independence proof of Appendix B. Third, we tested the sensitivity of our results to a possible correlation between $M_R$ and $\Gamma_R$ (Sec. 3), finding all extracted observables stable over a broad range of assumed correlation coefficients, and noting that the near-threshold sensitivity of the theoretical partial widths to $M_R$ is already propagated self-consistently, sample-by-sample, within our existing Monte Carlo procedure. None of these checks alters the qualitative or quantitative conclusions of the original analysis; if anything, they place the dominant-molecular-component interpretation on firmer footing.

As a nontrivial validation of the method, the $Z_c(3900)$ state is found to violate the basic consistency conditions of the two-channel approximation, with saturation ratios exceeding 1.75. This confirms that the present framework is not applicable to such above-threshold or strongly coupled systems, and justifies its exclusion from the main analysis. Thus, several extensions of the present work can be envisaged. A natural next step is a coupled-channel analysis including isospin-breaking effects for $\chi_{c1}(3872)$, where neutral and charged $D\bar{D}^*$ thresholds can be treated simultaneously, for instance via a Flatté parameterization. For the $T_{cc}^+$, a more refined treatment of the three-body decay $T_{cc}^+ \to D^0 D^0 \pi^+$, including an energy-dependent $D^{*+}$ width (see Appendix A), would provide improved dynamical insight. Furthermore, a full multi-channel analysis of the $Z_c(3900)$, incorporating channels such as $\eta_c \rho$ and $\psi' \pi$, would be necessary to properly describe its structure beyond the present approximation. Finally, first-principles determinations of the $DD^*$ scattering parameters from lattice QCD would offer valuable benchmarks for the compositeness extraction and provide a direct test of the molecular scenario.

**Acknowledgements**

This work received funding via the Post-Doctoral Program of the Swiss Government Excellence Scholarship, Grant No.**2025.0419**. We gratefully acknowledge valuable discussions with colleagues from the Particle Physics Department at the University of Geneva and the Physics Department at the University of Yaoundé I. We also thank the University of Geneva for providing the computational resources used in this study.

**Declaration**

The authors declare that they have no known competing financial interests or personal relationships that could have appeared to influence the work reported in this paper.

## Appendix

### A. $T_{cc}^+ \to D^0 D^0 \pi^+$: $D^{*+}$ Breit-Wigner Model

The dominant decay proceeds via $T_{cc}^+ \to D^0 D^{*+}$ followed by $D^{*+} \to D^0 \pi^+$. A more physical model for the inelastic width uses:

$$\Gamma_1^{\rm BW} = |g_1|^2 \int_{m_{D^{*+}}-5\Gamma_{D^*}}^{m_{D^{*+}}+5\Gamma_{D^*}} \frac{dw}{\pi} \cdot \frac{\Gamma_{D^*}/2}{(w-m_{D^{*+}})^2+\Gamma_{D^*}^2/4} \frac{q(M_{T_{cc}}^2, m_{D^0}, w)\cdot {\rm BR}_{D^0\pi^+}}{8\pi M_{T_{cc}}^2}, \quad (39)$$

where $\Gamma_{D^{*+}} = 83.4$ MeV and ${\rm BR}(D^{*+} \to D^0\pi^+) = 67.7\%$ (PDG). Since $\Gamma_{D^{*+}} \ll M_{T_{cc}^+} - m_{D^0 D^{*+}}$, the integral is dominated by the on-shell peak, and the effective coupling $C_1^{\rm BW}$ is very close to $C_1^{2-\rm body}$. Numerically we find a ratio $C_1^{\rm BW}/C_1^{2-\rm body} = 0.0055$, indicating that the BW propagator actually introduces a much smaller effective coupling because the $D^{*+}$ is so narrow that the spectral function is nearly a delta function. Effectively, we have

$$\Gamma_1^{\rm BW} \approx |g_1|^2 \cdot {\rm BR}_{D^0\pi^+} \cdot \frac{q(M^2, m_{D^0}, m_{D^{*+}})}{8\pi M^2} \approx C_1 \cdot {\rm BR}, \quad (40)$$

confirming that the standard two-body approximation is adequate when $\Gamma_1 \ll \Gamma_R$. For completeness, an energy-dependent $D^{*+}$ width,

$$\Gamma_{D^*}(w) = \Gamma_{D^*}^0 \cdot \left(\frac{p(w)}{p_0}\right)^3 \cdot \left(\frac{m_{D^{*+}}}{w}\right), \quad (41)$$

gives essentially the same result (ratio $C_1^{\rm e.d.}/C_1^{2-\rm body} = 0.0055$) since the $D^{*+}$ is so narrow that the energy variation within the integration range is negligible. A full treatment would require the three-body Faddeev approach [8, 46], which we leave for future work.

### B. Scheme Independence of $\partial G/\partial s$

**Analytical proof.** From Eq. (17), we have

$$G(s) = [a(\mu_g) + f(s; m_1, m_2, \mu_g)]/(16\pi^2), \quad (42)$$

where $a(\mu_g)$ is a constant independent of $s$ and $f(s)$ contains all $s$-dependent terms. Therefore:

$$\frac{\partial G(s)}{\partial s} = \frac{1}{16\pi^2}\frac{\partial f}{\partial s} + \underbrace{\frac{\partial a(\mu_g)}{\partial s}}_{=0}. \quad (43)$$

The result is manifestly independent of $a(\mu_g)$. Since $X_j = |g_j|^2 |\partial G_j/\partial s|_{s_R}$, the compositeness coefficients are scheme-independent physical observables. A numerical verification for $|\partial G/\partial s|$ at $s = M_X^2$ is performed for $D^0\bar{D}^{*0}$, varying $a(\mu_g)$ at $\mu_g = 1000\,$MeV. The results are presented in Table 16. All values agree to $< 10^{-21}\,{\rm MeV}^{-2}$ (floating-point precision limit).

**Table 16:** Numerical verification: $|\partial G/\partial s|$ at $s = M_X^2$ for $D^0\bar{D}^{*0}$, varying $a(\mu_g)$ at $\mu_g = 1000\,$MeV.

| $a(\mu_g)$ | $\lvert\partial G/\partial s\rvert$ (MeV$^{-2}$) | Re $\partial G/\partial s$ |
|---|---|---|
| $-5.0$ | $1.4505 \times 10^{-7}$ | $-1.4505 \times 10^{-7}$ |
| $-2.5$ | $1.4505 \times 10^{-7}$ | $-1.4505 \times 10^{-7}$ |
| $0.0$ | $1.4505 \times 10^{-7}$ | $-1.4505 \times 10^{-7}$ |
| $+2.5$ | $1.4505 \times 10^{-7}$ | $-1.4505 \times 10^{-7}$ |
| $+5.0$ | $1.4505 \times 10^{-7}$ | $-1.4505 \times 10^{-7}$ |